\documentstyle[sprocl,epsf]{article}

\def\bfsigma{\mbox{\boldmath $\sigma$}}

\begin{document}
\rightline{HEPHY-PUB 696/98}
\rightline{UWThPh-1998-33}\vskip 1truecm

\title{QUARK CONFINEMENT AND THE HADRON SPECTRUM }

\author{Nora Brambilla\footnote{Lectures given by Nora Brambilla at HUGS at CEBAF, $13^{th}$ Annual Hampton 
 University Graduate Studies at the Continuous Electron Beam Facility, May 26-June 12, 1998.}}
\address{Institute for  Theoretical  Physics, University Vienna \\ Boltzmanngasse 5, A-1090 Vienna, Austria 
        \\E-mail: brambill@doppler.thp.univie.ac.at   nora.brambilla@mi.infn.it}
\author{Antonio Vairo}
\address{High Energy Institute, Austrian Academy of Science 
        \\Nikolsdorfergasse  18, A-1050 Vienna, Austria
        \\E-mail: vairo@hephy.oeaw.ac.at  antonio.vairo@cern.ch}
\maketitle
\centerline{\it Dedicated to the 60th birthday of Dieter Gromes}
\vskip 0.4truecm
\abstracts{These lectures contain an introduction to the following topics:
\begin{itemize}
\item{} Phenomenology of the hadron spectrum.
\item{} The static Wilson loop in perturbative and in lattice QCD. Confinement and the flux tube formation. 
\item{} Non static properties: effective field theories and relativistic corrections to the quarkonium potential. 
\item{} The QCD vacuum: minimal area law, Abelian projection and dual Meissner effect, stochastic vacuum. 
\end{itemize}
}

\section{Introduction}
Quarks appear to be confined in nature. This means that free quarks have not been detected so far 
but only hadrons, their bound states. Therefore, predictions of the hadron spectrum are an explicit and direct 
test of our understanding of the confinement mechanism as a result of the low energy dynamics of QCD. 
In these lectures we will focus on heavy quark bound states and for simplicity we will only treat 
the quark-antiquark case i.e. the mesons. Indeed, also in order to extract the  Cabibbo--Kobayashi--Maskawa 
matrix elements and to study CP violation from the experimental decay rate of heavy mesons, hadronic matrix elements 
are needed. It is reasonable to expect that to this aim the somewhat simpler matching of the theoretical 
prediction for the  spectrum to experiment has to be previously established. 
Moreover, we need to achieve some understanding of the bound state dynamics in QCD 
in order to make reliable identifications of the gluonic degrees of freedom in the 
spectrum (hybrids, glueballs). All this is relevant to the programme of most of  
the accelerators machines, Godfrey et al. (1998). From a more general point of view, the issue about the 
consequences of a nontrivial vacuum structure and the nonperturbative definition of a field theory 
are questions that overlap with the domain of supersymmetric and string theories.

These lectures are quite pedagogical and introductory and contain several illustrative exercises. 
The interested reader is referred for details to the quoted references.

The plan of the lectures is the following one. In Secs. 2 and 3  we give a brief overview 
on the hadron spectrum and on the phenomenological models devised to explain it.
The nonperturbative phenomenological parameter $\sigma$ is introduced and connected 
to the Regge trajectories as well as to the string models. In Sec. 4  we evaluate 
perturbatively and nonperturbatively (via lattice simulations) the static Wilson loop.  
We discuss the area law behaviour and the flux tube formation as signals of confinement. 
In Sec. 5 we summarize some existing model independent results on the heavy quark interaction.
QCD effective field theories are introduced. In Sec. 6 we connect confinement to 
the structure of the QCD vacuum. We study with some detail the Minimal Area Law model. 
We discuss the Abelian Higgs model and the Dual Meissner effect and introduce the idea of 't Hooft 
Abelian projection. We list some of the results obtained on the lattice with partial gauge fixing.  
Finally we briefly review two models of the QCD vacuum: Dual QCD and the Stochastic Vacuum Model. 
Each section is supplemented with some exercises. We tried to be as self-contained as possible reporting 
all the relevant definitions and the basic concepts.

\section{The Hadron Spectrum}
The meson and the baryon resonances together with an introduction to the quark model have 
been discussed at this school by Jim Napolitano. Since these lectures are mainly concerned with 
the quark confinement mechanism, they contain only a general overview  on the spectrum
pointing out its relevant features. 

\begin{figure}[thb]
\makebox[3.0truecm]{\phantom b}
\epsfxsize=5truecm
\epsffile{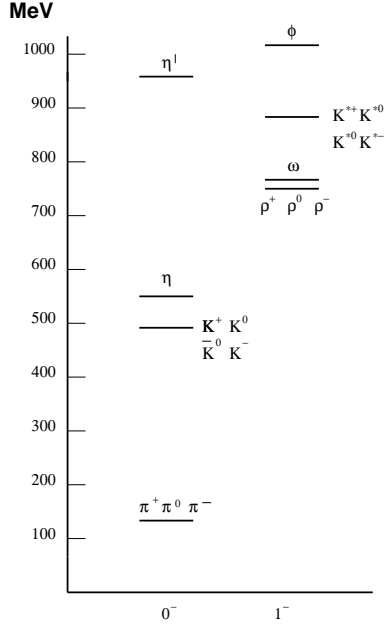}
\caption{\it The spectrum of the lightest mesons labeled by $({\rm spin})^{\rm parity}$.} 
\label{pluno}
\end{figure}

\begin{figure}[thb]
\makebox[2.0truecm]{\phantom b}
\epsfxsize=8truecm\epsffile{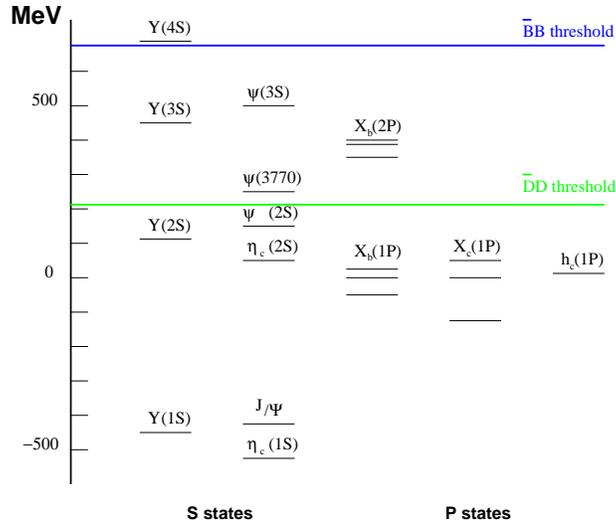} 
\caption{\it The experimental heavy  meson spectrum ($b\bar{b}$ and $c\bar{c}$) relative to the 
spin-average of the $\chi_b(1P)$ and $\chi_c(1P) $ states.} 
\label{pldue}
\end{figure}

\begin{figure}[thb]
\vskip -2truecm
\makebox[1.0truecm]{\phantom b}
\centerline{
\epsfxsize=8truecm
\epsffile{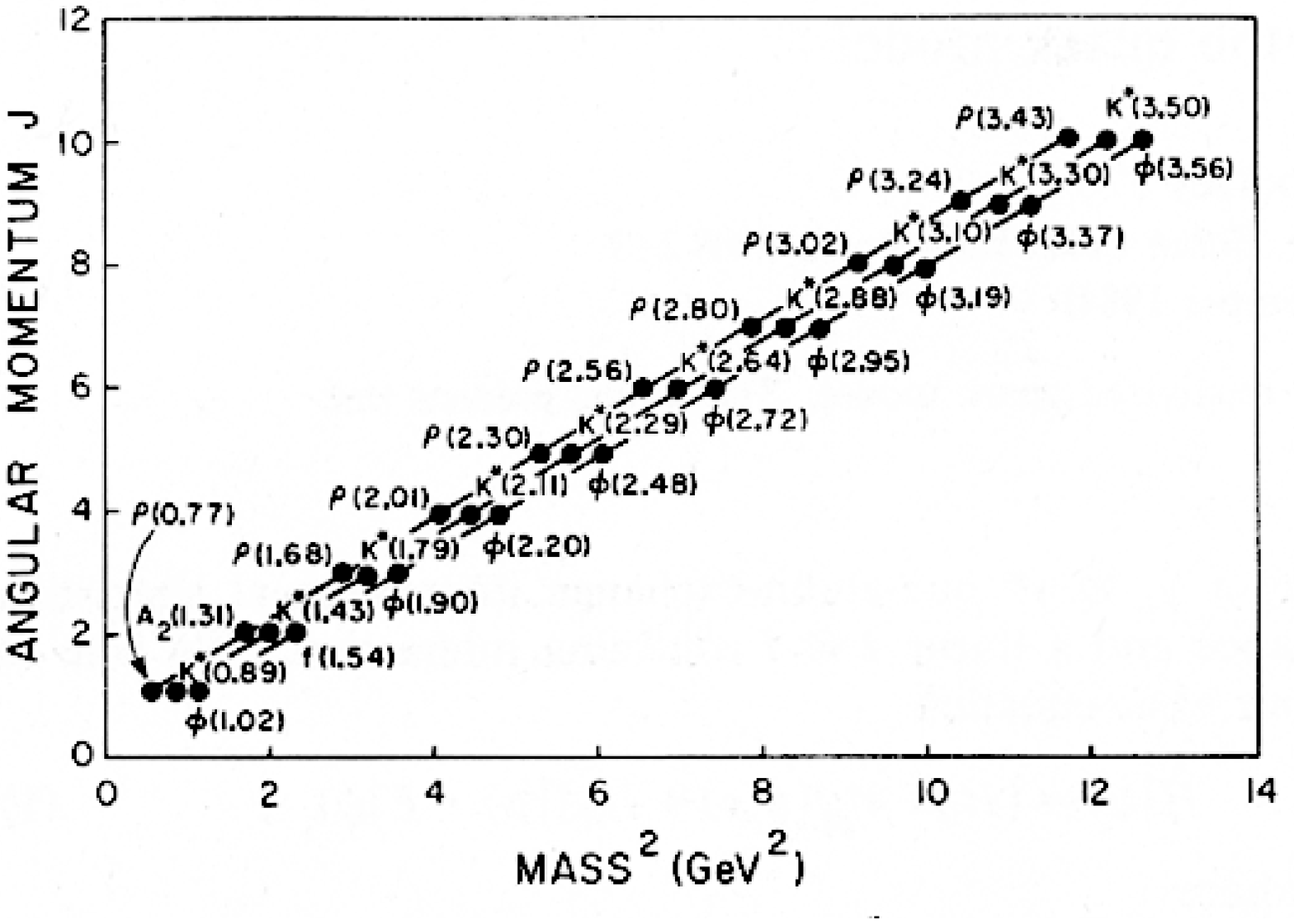}}
\vskip -3truecm
\caption{\it The Regge trajectories for the $\rho$, $K^*$ and $\phi$. From Godfrey et al. (1985).} 
\label{pltre}
\end{figure}

Let us concentrate on the meson spectrum as given in Figs. \ref{pluno}-\ref{pltre}.
In principle, one should be able to explain and predict it only by means of the QCD Lagrangian
\begin{equation}
L= -{1\over 4} F^{(a)}_{\mu\nu} F^{(a)\mu \nu} + i\sum_{q=1}^{N_f} \bar{\psi}_q^i\gamma^\mu (D_\mu)_{ij}\psi_q^j 
- \sum_{q=1}^{N_f} m_q \bar{\psi}_q^i\psi_q^i
\label{lagr}
\end{equation}
where 
\begin{eqnarray}
F_{\mu \nu} & & \equiv  \partial_\mu A_\nu-\partial_\nu A_\mu +ig [A_\mu, A_\nu] \nonumber \\
(D_\mu)_{ij} 
& & \equiv  \delta_{ij}\partial_\mu + ig {\lambda^a_{ij}\over 2}A^a_\mu \, ; 
\quad \quad \qquad A_\mu\equiv A_\mu^a {\lambda_a\over 2} \quad  a=1, \dots ,8.    
\label{def}
\end{eqnarray}
$\psi^j_q$ are the quark fields of flavour $q=1,\dots, 6=N_f$ and colour $j =1,\dots , 3$, 
$m_q$  the (current) quark masses and $\lambda^a $ the Gell-Mann matrices of $SU(3)$.  
However, if we proceed to calculate the spectrum from this Lagrangian using the familiar tool of a   
perturbative expansion in the coupling constant $g$, we get no match with the experimental 
data presented in Figs. \ref{pluno}-\ref{pltre}. This is a consequence of the most relevant feature 
of the Lagrangian (\ref{lagr}): asymptotic freedom (Gross and Wilczek (1973), Politzer (1973)). 
The coupling constant ($\alpha_{\rm s} \equiv g^2/4 \pi$) vanishes in the infinitely high energy region 
and grows uncontrolled in the infrared energy region:
\begin{equation}
{d \, \alpha_{\rm s}(\mu) \over d \log \mu^2}\equiv \beta(\alpha_{\rm s}) = -\alpha_{\rm s} 
\left( \beta_0 {\alpha_{\rm s}\over 4\pi} + \beta_1 \left({\alpha_{\rm s}\over 4\pi}\right)^2 + \cdots \right)
\label{betaqcd}
\end{equation}
where $\beta_0 = 11 - 2/3 N_f > 0$. As a consequence, a perturbative treatment is expected 
to be reliable in QCD only when the energy $\mu$  is large compared to 
$\Lambda_{QCD}$, which is the infrared energy scale defined by Eq. (\ref{betaqcd})\footnote{
$\Lambda_{\rm \overline{MS}}^{N_f=4}= 305\pm 25\pm 50$ MeV; this value corresponds 
to $\alpha_s(M_Z)= 0.117 \pm 0.002\pm 0.004 $, Particle Data Group (PDG) (1998).}. 
The point is that the relevant energies in a quark bound state are in most cases 
of the order of the naturally occurring scale of 1 fm ($\simeq \Lambda_{QCD}^{-1}$), the average hadron size. 
The only exception are the mesons made up with top quarks. Unfortunately, they have not enough time to exist!\footnote{
The top quark decays into a real $W$ and a $b$ with a large width. The toponium life-time would be even 
smaller than the revolution time, thereby precluding the formation of a mesonic bound state, Quigg (1997), 
Bigi et al. (1986).} 

The masses appearing in the Lagrangian are the so-called ``current mas\-ses'' and fall into two categories:
light quark masses $m_u=1.5 \div  5$ MeV, $m_d= 3 \div 9$ MeV, $m_s= 60 \div 170$ MeV
($m_u$, $m_d\ll \Lambda_{QCD}$ and $m_s \sim \Lambda_{QCD}$) and 
heavy quark masses $m_c=1.1 \div 1.4$ GeV, $m_b= 4.1 \div 4.4$ GeV, $m_t=173.8\pm 5.2$ GeV 
($m_c, m_b, m_t \gg \Lambda_{QCD}$)\footnote{The masses are scale dependent objects.
For what concerns the experimental values quoted above (PDG (1998)),
the $u$, $d$, $s$ quark masses are estimates of the so-called current quark masses in a mass independent 
subtraction scheme such as $\overline{\rm MS}$ at a scale $\mu \simeq 2$ GeV. The $c$ and $b$ quark masses 
are estimated from charmonium, bottomonium, $D$ and $B$ masses. They are the running masses in the 
$\overline{\rm MS}$ scheme. These can be different  from the constituent masses obtained in potential models,
see below. We remark that it exists a definition of the mass, the pole quark mass (appropriate 
only for very heavy quarks) which does not depend of the renormalization scale $\mu$.
The quark masses are calculated in lattice QCD, QCD sum rules, chiral perturbation theory. 
For further explanations see Dosch and Narison (1998), Jamin et al. (1998), Kenway (1998), Leutwyler (1996).}. 
The spectrum should be obtained from the QCD Lagrangian with these values of the masses. 
However, in the light quark sector, spontaneous chiral symmetry breaking and non-linear strongly 
coupled effects cooperate in a highly non trivial way. Indeed, it is peculiar of  QCD that, due 
to confinement, the quark masses are not physical, i.e. directly measurable quantities. 
Therefore, it turns out to be useful, in order to make phenomenological predictions, to introduce the so-called 
constituent quark masses, containing the current masses as well as mass corrections also due to confinement effects. 
These constituent masses can be defined, for example, using the additivity of the quark magnetic moments 
inside a hadron or using phenomenological potential models to fit the spectrum. For quarks heavier than  
$\Lambda_{QCD}$ the difference between current and constituent masses is not quite relevant.

In the framework of the constituent quark model the meson states are classified as follows. 
For equal masses the quark and the antiquark spins combine to give the total spin 
${\bf S}={\bf S}_1 + {\bf S}_2$ which combines with the orbital angular momentum ${ \bf L}$ to give 
the total angular momentum ${ \bf J}$. The resulting state is denoted by $n ^{2 S+1}L_J$ where 
$n-1$ is the number of radial nodes.  As usual, to $L=0$ is given the name $S$, to $L=1$ the name $P$, 
to $L=2$ the name D and so on. The resonances are  classified via the $J^{PC}$ quantum numbers, 
$P= (-1)^{L+1} $ being  the parity number and $C= (-1)^{L+S}$  the C-parity.

\begin{figure}[t]
\vskip -1.5truecm
\makebox[3.0truecm]{\phantom b}
\epsfxsize=6truecm
\epsffile{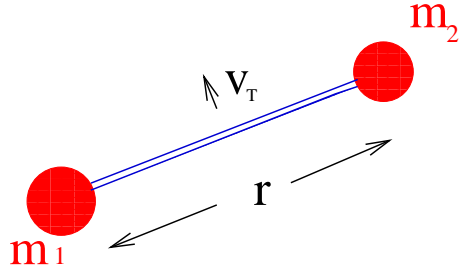}
\caption{\it Picture of the quark-antiquark bound state in the string model.} 
\label{pltube}
\end{figure}

Light mesons (as well as baryons) of a given internal symmetry quantum number but with different spins 
obey a simple spin ($J$)-mass ($M$) relation. They lie on a Regge trajectory
\begin{equation}
J(M^2)=\alpha_0 +\alpha^\prime M^2
\label{reg}
\end{equation}
with $\alpha^\prime \simeq 0.8 -0.9 \,  {\rm GeV}^{-2}$, see Fig. \ref{pltre}.
Up to now, free quarks have not been detected. The upper limit on the cosmic abundance
of relic quarks, $n_q$, is $n_q/n_p< 10^{-27}$, $n_p$ being the abundance of nucleons, while 
cosmological models predict $n_q/n_p< 10^{-12}$ for unconfined quarks.
The fact that no free quarks have been ever detected hints to the property of quark confinement. 
Hence, the interaction among quarks has to be so strong at large distances that a 
$q\bar{q}$ pair is always created when the quarks are widely separated. 
From the data it is reasonable to expect that a quark typically comes accompanied by an antiquark 
in a hadron of mass $ 1 \, {\rm GeV}$ at a separation of 1 {\rm fm} ($\simeq \Lambda_{QCD}^{-1}$). 
This suggests that between the  quark and  the antiquark there is a linear energy density (called string 
tension) of order 
\begin{equation}
\sigma = {\Delta E\over \Delta r} \simeq 1 {{\rm GeV}\over {\rm fm}}\simeq 0.2 \, {\rm GeV}^2 .
\label{sigma1}
\end{equation}
The evidence for linear Regge trajectories (see Fig. \ref{pltre}) supports this picture. 
A theoretical framework  is provided by the string model, Nambu (1974). In this model the hadron 
is represented as a rotating string with the two quarks at the ends. The string is formed
by the chromoelectric field responsible for  the flux tube configuration and for the quark confinement
(see Fig. \ref{pltube}), Buchm\"uller (1982). Upon solution of the Exercise 2.1, the reader can verify that 
it is possible to establish the relation
\begin{equation}
\alpha^\prime = {1\over 2 \pi \sigma }
\label{rel}
\end{equation}
between the slope of the Regge trajectories and the string tension.  The string tension $\sigma$ emerges 
as a key phenomenological parameter of the  confinement physics.

From the light mesons spectrum of Fig. \ref{pluno} it is evident that 
\begin{itemize}
\item{} the separations between levels are  considerably larger than 
the mesons masses $\Longrightarrow$ they are truly relativistic bound states;
\item{}  the splitting between the pseudoscalar $\pi$ and the vector $\rho$ mesons 
is so large to be anomalous $\Longrightarrow$  it  is due to the Goldstone boson nature of the $\pi$.
\end{itemize}
Therefore,  understanding  the light meson spectrum means to solve  a relativistic many-body 
bound state problem where confinement is strongly related with the spontaneous breaking 
of chiral symmetry. Since the target of these lectures is to gain some understanding of the confinement 
mechanism in relation to the spectrum, we will try to separate problems and to consider first the spectrum 
of mesons built by heavy valence quarks only. In this case we have still  bound states of confined  quarks, 
however, due to the large  mass of the quarks involved, we can hope to treat relativistic 
and many-body (i.e. quark pair creation effects) contributions as corrections. 
The problem simplifies remarkably if we consider mesons made up by two valence heavy quarks 
($b\bar{b}$, $c\bar{c}$, $b\bar{c}$, ...) i.e. quarkonium.

\vskip 1truecm
\leftline{\bf 2. Exercises}
\begin{itemize}
\item[2.1]{Consider two massless and spinless quarks connected by a string of length $R$ rotating 
with the endpoints at the  speed of light, so that each point at distance $r$ from the centre has the local 
velocity $v/c = 2 r/R$.  Recalling that the string tension $\sigma$ is the linear energy 
density of the string between the quarks, calculate the total mass and the total angular 
momentum and demonstrate that they lie on a Regge trajectory with $\alpha^\prime = 1 / (2 \pi \sigma)$.}
\item[2.2]{Consider the spin-independent Lagrangian
\begin{eqnarray}
L & & = - m_1 \sqrt{1-{\bf v}^2_1}- m_2 \sqrt{1-{\bf v}^2_2}-U_0(r)\nonumber \\
& & -{U_+(r)\over 4 } ({\bf v}_1 -{\bf v}_2)^2
-{U_-(r)\over 4 }({\bf v}_1 +{\bf v}_2)^2
\nonumber
\end{eqnarray}
where  ${\bf x}_1, {\bf v}_1$ and $ {\bf x}_2, {\bf v}_2 $ are the positions  and velocities respectively, 
of the quark and the antiquark and ${\bf r}={\bf x}_1 -{\bf x}_2$. $U_0$ is the static potential 
and $U_+ $ and $U_-$ are the coefficients of the velocity dependent terms in the potential.
Velocity dependent terms of this type are obtained in Baker et al. (1995), Brambilla et al. (1994). 
Making some simplifications (circular  orbits, ${v}_j\equiv {\bf \omega} \times {\bf r}_j$ 
with ${\bf r}_j=(-1)^{j+1} r_j \hat{{\bf r}}$, $r_1+r_2=r$, $m_1=m_2$), 
calculate the energy $E$ and the angular momentum ${\bf J}$. Determine the 
moment of inertia of the colour field produced by the rotating quarks and establish the physical 
meaning of $U_+$. Then, obtain the slope of the Regge trajectories in the case in which 
$U_0=\sigma r$ and $U_+ = -A r$.}
\end{itemize}

\section{Quarkonium and Confining Phenomenological Potentials}
The relevant features of the quarkonium spectrum are: the pattern of the  levels, the spin separation
between pseudoscalar mesons $n^1S_0 (0^{-+})$  and vector mesons  $n^3S_1 (1^{--})$ (called hyperfine 
splitting), the spin separations between states  within the same  $L\neq 0$ and $S$ multiplets 
(e.g. the splitting in the $ 1^3P_J$ multiplet $\chi_c(1P)$ in charmonium cf. Fig. \ref{pldue}) 
(called fine splitting), and the transition and decay rates, see PDG. 
To separate the sub-structure from the radial and orbital splittings it is convenient 
to work in terms of spin-averaged splittings.
Spin-averaged states are obtained summing over masses of given $L$ and $n$, and weighting by $2 J+1$.
The hyperfine spin splittings appear to scale roughly with a $1/m_Q$ dependence.
We note that states below threshold are considerably narrow since they can decay only by annihilation.

The fact that all the splittings are considerably smaller than the masses implies that all the dynamical
scales of the bound state, such  as the kinetic energy or the momentum of the heavy quarks, 
are considerably smaller than the quark masses. Therefore, the quark velocities are nonrelativistic: $v\ll 1$.
The energy scales in quarkonium are the typical scales of a nonrelativistic bound state: 
the momentum scale  $m_Q v$ and the  energy scale $m_Q v^2$. Being the time scale $T_g \sim 1/m_Q v$ 
associated with the binding gluons smaller than the time scale $T_Q \sim 1/m_Q v^2$ associated with 
the quark motion, the gluon interaction between heavy quarks appears ``instantaneous''. 
Therefore, it can be modelled with a potential and the energy can be obtained solving the corresponding 
Schr\"odinger equation. In the extreme nonrelativistic limit of very heavy quarks the spin splittings vanish and 
the spin-averaged spectrum is described by a single static central potential.

\begin{figure}[htb]
\vskip -0.2truecm
\makebox[1.0truecm]{\phantom b}
\centerline{ 
\epsfxsize=8truecm
\epsffile{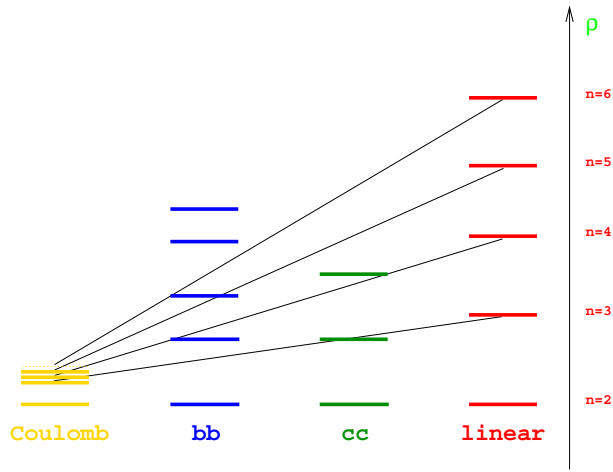}}
\vskip -0.2truecm
\caption{\it $\rho = (E_n- E_1)/(E_2-E_1)$, where $E_n$ is either the $n$th energy level of the physical 
system or the $n$th eigenvalue of the Schr\"odinger equation corresponding to the indicated potential.} 
\label{pllevel}
\end{figure}

A lot of work has been done to find  the phenomenological form of  the static potential (see the report 
of Gromes, Lucha and Sch\"oberl (1991)). Lowest order perturbation theory for QCD gives a flavour-independent 
central potential based on the one-gluon exchange, which has a Coulomb-like form (see Exercise 3.1)
\begin{equation}
V_0(r)= - {4\over 3} {\alpha_{\rm s} \over r}
\label{coulomb}
\end{equation}
where $r$ is the distance between the two quarks and $\alpha_{\rm s}$ is the strong coupling constant
\footnote{To compute Eq. (\ref{coulomb}), cf. Exercise 3.1, it is necessary to evaluate products 
of the type $\lambda^{(1)}\cdot \lambda^{(2)}$ in the representation of interest. 
It turns out that, out of all two-body channels, the colour singlet ($q\bar{q}$) is the most attractive. 
This hints to the fact  that coloured mesons should not exist.}
\footnote{Of course, $\alpha_{\rm s}$ in Eq. (\ref{coulomb}) depends on a scale. If we work
in a physical gauge, e.g. in a lightlike gauge, the quarks in the bound state interact through 
gluon exchange and these interactions are renormalized by loop corrections. 
In the extreme nonrelativistic limit, only the corrections to the gluon propagator survive and this
give $\alpha_{\rm s}$ evaluated at the square momentum of the exchanged gluon $Q^2\simeq {\bf Q}^2$.}. 
This cannot be the final answer since it does not confine quarks and  gives a spectrum 
incompatible with the data (see Fig. \ref{pllevel}). 
Nevertheless we can regard  the one-gluon exchange formula to be valid for $r \simeq 0.1$ fm. 

It was found that the addition of some positive power of $r$ to Eq. (\ref{coulomb})
rescues the phenomenology. The intuitive argument of a constant energy density (string tension), exposed
in Sec. 2,  led to the flavor-independent  Cornell potential (Eichten et al. (1978))
\begin{equation}
V_0(r) = -{4\over 3} {\alpha_{\rm s}\over r } +\sigma r + {\rm const.}
\label{cornell}
\end{equation}
Here, $\alpha_{\rm s}$ and $\sigma$ are regarded as free parameters to be fitted on the spectrum.
The  Schr\"odinger equation with the potential (\ref{cornell}) and parameters $\alpha_{\rm s} = 0.39$ 
and $\sigma=0.182$ GeV$^2$ gives  quite a satisfactory agreement with the data.
In Eichten et al. (1980), the coupling to charmed meson decay-channels was also taken into account. 
It was found that the mass shifts due to the coupled channel effects are indeed large also below  threshold 
and yet they do not spoil the predictions of the naive potential model. Indeed these effects can essentially  
be absorbed into  a redefinition of the effective parameters. 
\begin{figure}[htb]
\vskip -2truecm
\makebox[1.0truecm]{\phantom b}
\centerline{ 
\epsfxsize=8truecm
\epsffile{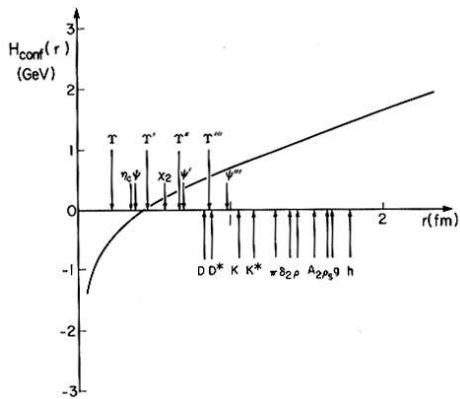}}
\vskip -2truecm
\caption{\it The rms $q\bar{q}$ separations in some representative mesons is shown with respect to 
the Cornell potential. All the phenomenological potentials agree in the range of $0.1-1$ fm which is 
the physical range for quarkonia. From Godfrey and Isgur (1985).} 
\label{plpot}
\end{figure}
Since then several different phenomenological forms of the static potential have been exploited,
e.g. the Richardson potential, $${\displaystyle V_0(r)= \int {d^3 {\bf Q}\over (2 \pi)^3} 
\exp{\{i {\bf Q}\cdot {\bf r}\}} {{\rm const.}\over Q^2\log(1+  Q^2/\Lambda^2)}}$$ 
(Richardson (1979)), the logarithmic potential, $V_0(r)= A\log (r/ r_0)$ (Quigg and Rosner (1977)), 
the Martin potential, $V_0(r)= A(r/ r_0)^\alpha$ (Martin (1980)).
By fitting the parameters, all  these potentials can reproduce the spectrum. This is not surprising 
if we look at Fig. \ref{plpot}: the potentials essentially  agree in the region $r \sim 0.1\div0.8$ fm 
in which the $\sqrt{\langle r^2\rangle}$ for quarkonia sits (Buchm\"uller and Tye (1981)). 
On the other hand, it is  a considerable limit of the potential approach 
that the connection with the true QCD  parameters of Eq. (\ref{lagr}) remains totally hidden and mysterious. 

However, the existence of a spin substructure in  the spectrum  indicates that relativistic corrections 
to the static central potential $V_0$ have to be taken into account.  From the radial level splitting 
(see Exercise 3.2) as well as from the fits with the phenomenological potential, we find for 
\begin{eqnarray}
c \quad {\rm in } \quad  \psi, & &  v^2 \simeq 0.3 \nonumber \\
b  \quad {\rm in }\quad   \Upsilon, & & v^2 \simeq 0.1 \quad , 
\label{modvel}
\end{eqnarray} 
and therefore we expect relativistic corrections of order $20\div 30 \%$ for the charmonium spectrum 
and up to $10 \%$ for the bottomonium spectrum. This determination of the heavy 
quark velocity in the bound states is confirmed by lattice calculation, see Tab.1.
Notice that relativistic corrections are of critical importance for the observables sensitive to the details 
of the wave functions (e.g. the radiative  transitions ) (Mc Clary et al. (1983)).
\begin{center}
\begin{table}
\begin{center}
\begin{tabular}{|c|c|c|c|c|}
\hline
$nL$&$\langle { v_b^2}\rangle$&$\langle{ v_c^2}\rangle$&
$\sqrt{\langle { r_b^2}\rangle}/$fm&$\sqrt{\langle { r_c^2}\rangle}/$fm\\\hline
$1S$&0.080&0.27&0.24&0.43\\
$2S$&0.081&0.35&0.51&0.85\\
$3S$&0.096&0.44&0.73&1.18\\
$1P$&0.068&0.29&0.41&0.67\\
$2P$&0.085&0.39&0.65&1.04\\
$1D$&0.075&0.34&0.54&0.87\\
\hline
\end{tabular}
\end{center}
\caption{\it From Bali et al. (1997). Notice that for a  Coulombic  system $ v \sim \alpha/  n$ and 
for a  confined   system $ v$ grows with $ n$.}
\label{pltab}
\end{table}
\end{center}

The phenomenological potential model predictions of the relativistic corrections are calculated by 
means of a Breit--Fermi Hamiltonian of the type
\begin{equation}
H= \sum_{j=1,2} \left(m_j + {p_j^2 \over 2 m_j} -{p_j^4 \over 8 m_j^3}\right) + V_0 + V_{\rm SD} + V_{\rm VD}.
\label{ham}
\end{equation}
The $1/m^2$  spin-dependent $V_{\rm SD} $ and velocity-dependent $V_{\rm VD} $ potentials 
are derived from  the semirelativistic reduction of the  Bethe--Salpeter (BS) equation for the quark-antiquark 
connected amputated Green function  or, equivalently at this level, from the  semirelativistic 
reduction of the quark-antiquark scattering amplitude with an effective exchange equal to the BS kernel.  
Several ambiguities are involved in this procedure, due on  one hand  to the fact that 
we do not know the relevant confining Bethe--Salpeter kernel, on the other hand  due to the fact 
that we have to get rid of the temporal (or energy $Q_0$, $Q=p_1-p_1^\prime$ being the momentum transfer) 
dependence of the kernel to recover a potential (instantaneous) description. It turns out that, 
at the level of the approximation involved, the spin-independent relativistic corrections at the order $1/ m^2$ 
depend  on the way in which $Q_0$ is fixed together with  the gauge choice of the kernel. 
The Lorentz structure of the kernel is also not known.  

On a phenomenological basis, the following ansatz for the kernel was intensively studied
\begin{eqnarray}
I(Q^2)= (2 \pi)^3 \left[ \gamma_1^\mu  \gamma_2^\nu P_{\mu\nu}  J_v(Q) + J_s(Q) \right]
\label{kernel}
\end{eqnarray}
in the instantaneous approximation $Q_0=0$. Notice that the effective  kernel above was taken with 
a pure dependence on the momentum transfer $Q$. But, of course, the dependence on the quark and antiquark 
momenta could have been more complicated. The vector kernel $\gamma_1^\mu  \gamma_2^\nu P_{\mu\nu}  
J_v(Q)$ corresponds to the one gluon exchange,  and e.g. in the Coulomb gauge $P_{\mu\nu}$  has the structure  
$P_{\mu\nu}= g_{\mu\nu}+\displaystyle{J_v^\prime(-{\bf Q}^2)\over J_v(-{\bf Q}^2)} Q_\mu Q_\nu
- \displaystyle{J_v^\prime(-{\bf Q}^2)\over J_v(-{\bf Q}^2)} Q_0(Q_\mu n_\nu+Q_\nu n_\mu)$ 
where the prime indicates the derivative and $n_\mu$ is the unit vector in the time direction. 
The scalar kernel $J_s(Q)$ accounts for the nonperturbative interaction.
The semirelativistic reduction of the kernel (\ref{kernel}) (in the instantaneous approximation, 
with the Coulomb gauge fixed for the vectorial part, in  the centre of 
mass frame ${\bf p}_1= -{\bf p}_2 ={\bf q}, {\bf p}^\prime_1= -{\bf p}^\prime_2 ={\bf q}^\prime, 
{\bf Q} \equiv {\bf q} -{\bf q}^\prime$ and in the equal mass case), gives
\begin{eqnarray}
V_0    &=& \tilde{J}_s(r) + \tilde{J}_v(r) \label{vfenum}\\
V_{\rm SD} &=&  {3\over m^2} {1\over r}
 \left({d\tilde{J}_v\over dr} -{1\over 3} {d\tilde{J}_s\over dr}\right) {\bf S} \cdot {\bf L}
\label{vspin} \\
&+&{1\over m^2} \left({1\over r} {d\tilde{J}_v \over dr} - {d^2 \tilde{J}_v\over dr^2}\right) S_1^h
\left({r^hr^k\over r^2}-{\delta^{hk}\over 3}\right) S^k_2 + {2\over 3 m^2} \Delta \tilde{J}_v(r) 
{\bf S}_1 \cdot {\bf S}_2
\nonumber \\
V_{\rm VD} &=& {1\over 4 m^2 } \Delta(\tilde{J}_s +\tilde{J}_v) -{1\over 4 m^2} q^h \tilde{J}_s q^h 
+{1\over 2 m^2} q^h \left(\delta^{hk} -{\partial_h \partial_k \over \partial^2} \tilde{J}_v\right) q^k  
\label{vvel}
\end{eqnarray}
with $\tilde{J}_{v,s}({\bf r}) \equiv \displaystyle\int {d^3 {\bf Q}}  
\, e^{i{\bf Q} \cdot {\bf r} } J({\bf Q})_{v,s}$. Taking $J_v \! = \!\displaystyle-{1\over 2 \pi^2} {4\over 3} 
{1\over {\bf Q}^2}$ and $J_s \! = \! \displaystyle -{\sigma\over \pi^2} {1\over {\bf Q}^4}$,  
$V_0$ reproduces the Cornell potential.  

The confining part of the kernel is usually chosen to be a Lorentz scalar in order to match the data 
on the fine separation. Indeed, the ratio $\rho_{FS}$ of the fine structure splitting 
\begin{equation}
\rho_{FS} = {M( ^3P_2)- M( ^3P_1)\over M( ^3P_1)-M( ^3P_0)}
\end{equation}
is $\rho_{FS}= 0.8 $ for a pure vector Coulomb exchange while the data give $\rho_{FS}
\simeq  0.49, 0.66, 0.57$
for the $c\bar{c}(1P), b \bar{b}(1P)$ and $ b\bar{b}(2P)$ respectively. Adding a scalar exchange 
gives a contribution to the spin orbit interaction that reduces the vector part and thus also the 
value of the ratio $\rho_{FS}$ (Schnitzer (1978)). Moreover, since  the two terms, vectorial short range and scalar 
long range exchange, contribute with opposite signs, we expect  that at high orbital excitations, where the $Q\bar{Q}$ 
pair probes large average distances (see Fig. \ref{plpot}), the triplet multiplet  inverts 
with reference to the ordering of the low excitation multiplets 
(i.e.  $M( ^3L_{L+1})\ge M( ^3L_{L}) \ge M( ^3L_{L-1})$ at low $L$ and the reverse 
at high $L$).  The fine structure turns out to be a nice test for the form of the confining interaction
 (cf. e.g. Isgur (1998)).

We have presented a way to obtain phenomenologically the ${1/ m^2}$ relativistic corrections.
However, this is not really rewarding. In a confining interaction the  average $\langle p^2\rangle $ 
increases with the excited states (see Tab. 1) and so the pattern of the excited levels is likely to be considerably  
distorted if the nonperturbative relativistic corrections are  not the appropriate ones. Indeed, 
fits made with only the static potential turn out to be better than fits made with the 
interaction (\ref{vfenum})-(\ref{vvel}) (Brambilla et al. (1990)).   
Moreover, there are characteristics of the spectrum, like the hyperfine separation as well 
as the leptonic decays, that are due to processes taking place at very short 
scale. In this case it is important to add higher order  perturbative corrections to the one gluon exchange as well 
as the running of $\alpha_{\rm s}$. In the phenomenological potential framework it is somehow ambiguous 
how to take into account the running of  $\alpha_{\rm s}(\mu)$ as well as the scale $\mu$.
 
In this section we realized  that the description of the heavy meson spectrum turns out to be a priori 
a quite complicate problem with an interplay of different relevant scales as well as of  perturbative and 
nonperturbative effects. The conclusion is that we badly need, on one hand a framework in which relativistic 
as well as perturbative corrections to the quark-antiquark interaction can be evaluated unambiguously 
and systematically and, on the other hand a clear, well founded  and eventually  computable approach 
to the long range quark-antiquark interaction which is essentially nonperturbative.  We need this 
both at a concrete level, in order to make quantitative predictions in which the size of the neglected 
terms can be estimated, and both at a fundamental level, in order to use the spectrum to get some 
insight into the confinement mechanism. 

In the next section we address the problem of how to study quark confinement beyond phenomenological
models i.e. in a QCD based framework and we give a criterium that decides whether  a gauge theory 
is confined or not.

\vskip 1truecm
\leftline{\bf 3. Exercises}
\begin{itemize}
\item[3.1]{Consider the quark-antiquark scattering 
\begin{equation}
q_i(p_1,\sigma_1) + \bar{q}_j(p_2,\sigma_2) \to q_k(q_1,\tau_1) +\bar{q}_l(q_2,\tau_2)
\label{scat}
\end{equation}
where $i,j,\cdots =1,2,3$ label the colour indices. Remembering that 
the quarks in the meson are in a colour-singlet state and introducing the meson colour 
wavefunctions $\delta_{ij}/\sqrt{3}$, calculate the $T$-matrix element, extract the first 
contribution in the nonrelativistic limit and obtain, via Fourier transform, the perturbative 
one gluon exchange potential of Eq. (\ref{coulomb}). Show that the  contribution of the annihilation
graph vanishes.}
\item[3.2]{Consider the average excitation energy 
in charmonium and bottomonium (e.g. $M_\Upsilon^\prime -M_\Upsilon $). This should be of the order of 
the average kinetic energy $E\simeq mv^2$. Taking $m$ to be roughly  half of the ground state mass 
obtain the estimates (\ref{modvel}) for the quark velocities.} 
\item[3.3]{Consider the same scattering of (\ref{scat}) with the exchange given in Eq. (\ref{kernel}).
Obtain the result (\ref{vfenum})-(\ref{vvel}) by computing the scattering matrix element of this process, 
expanding up to the ${1/ m^2}$ order and Fourier transforming.} 
\item[3.4]{Consider the same scattering matrix of Exs. 3.1 and 3.3 but with a kernel of the type 
$\displaystyle I= \gamma^1_5 \gamma^2_5 V_p(Q) $. Show that there is no static potential 
in the nonrelativistic limit of the matrix element. Discuss the result in relation to the deuteron.}
\end{itemize}

\section{The Wilson Loop: Confinement and Flux Tube Formation}
The most powerful technique in order to extract nonperturbative information from QCD is the 
lattice gauge theory approach. This has been undoubtedly successful and rewarding 
and has produced over the last years an impressive amount of results. 
Yet, in spite of almost two decades of intensive efforts the  characteristics  of QCD associated with 
colour confinement are  still not understood. It is our belief  that  some insight in the mechanism 
of confinement cannot  be obtained without developing, in strict connection with lattice QCD, also 
analytic methods. In this way information coming from the lattice can be inserted inside analytic models 
or vice versa lattice calculations can be used in order to interpret analytic models. 

To this aim we need an unambiguous way of establishing a gauge invariant and systematic 
procedure to calculate the quark dynamics. In Sec. 5, we will show that this is feasible in the case of 
heavy quarks in which the whole dynamics can be reduced to few expectation values of chromoelectric 
and chromomagnetic  fields that can be calculated analytically (once a model for the QCD vacuum is assumed) 
or numerically on the lattice. Then, the comparison between the two results supply us with hints  about  
the mechanism of confinement.

The simplest manifestation of confinement in quenched QCD is the linear rising of the potential 
$V_0(r)$ between static colour sources in the fundamental representation. In this section 
we will show how this has been clearly proved and connected to the formation of a chromoelectric flux tube 
between the quarks. The question of the nature and  the origin of these nonperturbative field configurations 
will be addressed in Sec. 6.

\subsection{The QCD static potential and the Wilson loop}
Let us consider a locally gauge invariant quark-antiquark singlet state\footnote{The contribution of the string 
to the potential vanishes  in the limit $T\to \infty$, Eichten et al. (1981) and Brambilla et al. (1999).} 
\begin{equation}
\vert \phi_{\alpha \beta}^{lj} \rangle \equiv {\delta_{lj}\over \sqrt{3}}
\bar{\psi}^i_\alpha(x) U^{ik}(x,y,C) \psi^k_\beta(y) \vert 0\rangle
\label{statgaug}
\end{equation}
where $i,j,k,l$ are colour indices (that will be suppressed in the following), 
$\vert 0\rangle$ denotes the ground state and the Schwinger string line has the form
\begin{equation}
U(x,y;C) =P \exp \left\{ i g \int_y^x A_\mu(z) \, dz^\mu \right\} \,  ,
\label{string}
\end{equation}
where   $A_\mu$ is  the gauge potential of Eq. (\ref{def}), $g$  the QCD coupling constant, and the integral is 
extended along the path $C$. The operator $P$ denotes the path-ordering prescription\footnote{
Path ordering prescription means operatively that one has to decompose the path $C$ connecting $y$ with $x$ 
into infinitesimal pieces, then take the exponential along the infinitesimal pieces, expand at the first 
order and order the factors according to their appearance along the path.} which is  necessary due to the 
fact that $A_\mu$ are non-commuting matrices.

Remembering that under a $SU(3)$ gauge transformation ${\cal V}(\theta)$ $=$ $\exp({-i \theta})$ 
$\simeq 1-i\theta$, the gauge potential undergoes the transformation 
$A_\mu$ $\to$  $A_\mu$ $+$ $i [\theta, A_\mu]$ $+$ $g^{-1} \partial_\mu \theta$, we  
ob\-tain the tran\-sfor\-ma\-tion law of the string
\begin{equation}
U^\prime(x,y;C) =\exp{ \{i\theta(x)}\} U(x,y;C) \exp{\{-i \theta(y)\} }
\label{stringtrasf}
\end{equation}
and then it is clear that (\ref{statgaug}) is a gauge-invariant state. Actually, it is a colour singlet
and we are interested only in colour singlet being the only existing initial and final  states.

The quark-antiquark potential can be extracted from the quark-antiquark Green function. A simple 
example clarifies in which way. Let us consider the following two-particle Green function
\begin{equation}
G(T) = \langle \phi({\bf x},0) \vert \phi({\bf y}, T)\rangle =
\langle \phi({\bf x},0) \vert \exp{(-iH T)}\vert \phi({\bf y}, 0)\rangle . 
\label{green}
\end{equation}
Inserting a complete set of energy eigenstates $\psi_n$ with eigenvalues $E_n$ and making a 
Wick rotation we find
\begin{eqnarray}
G(-i T) & = & \sum_n \langle \phi({\bf x},0) \vert \psi_n \rangle \langle \psi_n \vert 
 \phi({\bf y}, 0)\rangle \exp{(-E_n T)} \nonumber \\
&\to &   \langle \phi({\bf x},0) \vert \psi_0 \rangle \langle \psi_0 \vert 
 \phi({\bf y}, 0)\rangle \exp{(-E_0 T)} \quad  {\rm for }\quad T\to \infty
\label{inf}
\end{eqnarray}
which gives the Feynman--Kac formula for the ground state energy
\begin{equation}
E_0 = - \lim_{T \to \infty} {\log G(-iT) \over T}.
\label{fey}
\end{equation}
The only condition for the validity of Eq. (\ref{fey}) is that the $\phi$ states have a non-vanishing 
component over the ground state. The same is still true for finite $T$ if the overlap with the 
ground state is not too small. This is precisely the way in which  hadron masses are computed on 
the lattice. Of course, many tricks are used in order to maximize the overlap with the ground state in consideration.
If the $\phi$ state denotes a state of two exactly static particles interacting at a distance $r$, then 
the ground state energy is a function of the particle separation, $E_0\equiv E_0(r)$, 
and gives the potential of the first adiabatic surface. With this in mind, we will perform in the remaining 
of this section an explicit evaluation of the quark-antiquark Green function for infinitely heavy 
quarks ($m_j\to \infty$) and for large temporal intervals ($T\to \infty$).

In the following we will be working in the Euclidean space taking advantage of the usual
relation between Euclidean position $x^E_\mu$, momentum $k^E_\mu$, field $A^E_\mu$, gamma matrices
 $\gamma^E$ and the corresponding quantities in Minkowski space
\begin{eqnarray}
& & t^E= it^M \qquad\quad~ x_i^E=x_i^M \nonumber \\ 
& & k_4^E = -i k_0^M \qquad\> k_i^E= k_i \nonumber \\    
& & A_4^E= -i A_0^M \qquad\!  A_i^E=A_i^M \nonumber \\
& & \gamma_4^E=\gamma_0 \qquad\qquad\! \gamma_i^E= -i \gamma^i .
\label{euclidep}
\end{eqnarray}

Let us assume  that at a time $t=0$ a quark and an antiquark are created and  that they interact 
while  propagating  for a time $t=T$ at which they are annihilated. 
Then ($x_j=({\bf x}_j, T), y_j=({\bf y}_j, 0)$)  
\begin{eqnarray}
& & G_{\beta_1\beta_2\alpha_1\alpha_2}(T)  \nonumber\\
& & = \langle 0 \vert \bar{\psi}_{\beta_2}({\bf y}_2, 0)U(y_2,y_1) \psi_{\beta_1}({\bf y}_1,0)\bar{\psi}_{\alpha_1}
({\bf x}_1, T) U(x_1,x_2)\psi_{\alpha_2}({\bf x}_2,T)\vert 0\rangle \nonumber \\
& & = {1\over Z}\int {\cal D}\psi {\cal D}\bar{\psi} {\cal D}A 
\, \bar{\psi}_{\beta_2}({\bf y}_2, 0)U(y_2,y_1)\psi_{\beta_1}({\bf y}_1,0) \nonumber \\
& &\qquad\qquad \times \bar{\psi}_{\alpha_1}({\bf x}_1, T) U(x_1,x_2)\psi_{\alpha_2}({\bf x}_2,T) 
e^{-\int L \,\,d^4x}
\label{green2}
\end{eqnarray}
where $L$ is the Euclidean version of Eq. (\ref{lagr}), $L=L_{YM} + L_F= {1\over 4} F_{\mu\nu} F_{\mu \nu} +\bar{\psi}
(\gamma_\mu D_\mu + m) \psi $. The indices $\alpha, \beta$ are spinor indices, while  the detailed structure 
of the colour indices is not displayed.
Since the action is quadratic in the quark fields, it is possible to perform the fermion integration
\begin{eqnarray} 
& & G_{\beta_1\beta_2\alpha_1\alpha_2}(T)  \nonumber \\
& & =  {1\over Z}\int  {\cal D}A \, \big( {\rm Tr} \{ S_{\alpha_2 \beta_2}(x_2,y_2; A) U(y_2,y_1)
 S_{\beta_1\alpha_1}(y_1,x_1;A) U(x_1,x_2) \}\nonumber \\
& & \qquad - {\rm Tr} \{ S_{\beta_1 \beta_2} (y_1,y_2;A) U(y_2,y_1) \} {\rm Tr}
 \{ S_{\alpha_2 \alpha_1}(x_2,x_1;A) U(x_1,x_2) \} \big)
\nonumber \\
& & \qquad \times \det{K(A)}  e^{-\int L_{YM} d^4x} 
\label{green3}
\end{eqnarray}
where the trace is over the colour indices and $K$ is the fermionic determinant of the matrix
$K_{\alpha x \beta y}(A) \equiv [\gamma_\mu D_\mu + m]_{\alpha \beta} \delta^4(x-y)$.
In the following, we will assume the quenched approximation\footnote{In 
perturbation theory the logarithm of this determinant is given by the sum of Feynman diagrams consisting 
of fermion loops with an arbitrary number of fields $A_\mu$ attached to it. In the limit $m\to \infty$ 
this determinant approaches a constant (infinite but canceled by a factor in $Z$) and then $\det K= 1 \> +$ 
corrections of order $O(1/m^n)$. Therefore for heavy quarks the quenched approximation $\det K=1$ makes sense.
On the contrary the fermionic determinant associated to light quarks  in principle cannot  be neglected.  
This determinant will eventually be responsible for  the breaking of the string between the quarks.}, $\det K=1$. 
The second term in Eq. (\ref{green3}) describes quark-antiquark annihilation and hence appears 
only for quarks of the same flavour. Since this effect is dominated by the perturbative two or three 
gluons exchange in the s channel we will not consider this term any more here. Then we obtain 
\begin{equation}
G(T) \simeq {1\over Z}\int  {\cal D}A \, {\rm Tr} \{ S(x_2,y_2; A) U(y_2,y_1) 
S(y_1,x_1;A) U(x_1,x_2) \} e^{-\int L_{YM} d^4x}.
\label{utile}
\end{equation}

In Eq. (\ref{utile}) $S(x,y;A)$ denotes the quark propagator in the presence of the gluon field $A_\mu$. 
It obeys the equation
\begin{equation}
(\gamma_\mu D_\mu + m)S(x,y;A)= \delta^4(x-y). 
\label{direq}
\end{equation}
This is in principle a system of coupled partial differential equations that cannot be solved in a closed form for an 
arbitrary $A_\mu$. However, we are interested in the limit $m\to \infty$. In this approximation (Wilson (1974), 
Brown and Weisberger (1979)) we can  replace $S$ by the static solution $S_0$ obtained dropping the spatial part 
of the gauge-covariant derivative in (\ref{direq}) while maintaining the time component. 
This approximation maintains the manifest gauge invariance. Then  we have 
\begin{equation}  
(\gamma_4 D_4  + m)S_0(x,y;A)= \delta^4(x-y) 
\label{eqstat}
\end{equation}
which is an ordinary differential equation solvable in a closed form. Indeed, we can get rid of the $A_4$ in 
the equation making the ansatz
\begin{equation}
S_0(x,y;A) = {\rm P}  \exp{\left\{ ig \int_{x_4}^{y_4} dt A_4({\bf x},t)\right\} } \hat{S}_0(x-y)
\label{ans}
\end{equation}
with $\hat{S}_0$ satisfying
\begin{equation}
(\gamma_4\partial_4 + m) \hat{S}(x-y)=\delta^4(x-y). 
\label{rideq}
\end{equation}
Therefore the solution has the form
\begin{eqnarray}
S_0(x,y;A) &=& \delta^3({\bf x}-{\bf y}) {\rm P} e^{ ig \int_{x_4}^{y_4} dt A_4({\bf x},t)} 
\left\{\theta(x_4-y_4) {1+\gamma_4\over 2} e^{ -m (x_4-y_4)  }\right.\nonumber \\
& & \qquad\qquad \left. + \theta(y_4-x_4) {1-\gamma_4\over 2} e^{- m (y_4-x_4) }  \right\}.
\label{soles}
\end{eqnarray}
This expression shows that the time evolution of a (infinitely) heavy quark field 
consists purely in the accumulation of phase determined by $A_4$ and the quark mass (cf. Ex. 4.1.1).
The spatial delta function says that the infinitely heavy quark cannot propagate in space\footnote{
For this solution the annihilation term in Eq. (\ref{green2}) does not give contribution
since the  ${\bf x} = {\bf y}$ condition is not satisfied.}. 

\begin{figure}[htb]
\vskip -0.1truecm
\makebox[4.0truecm]{\phantom b}
\epsfxsize=4truecm\epsffile{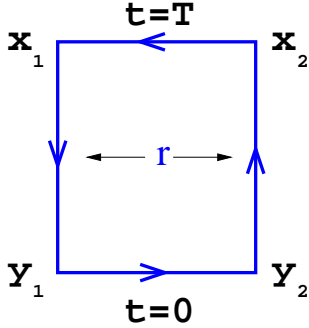}
\vskip 0.1truecm
\caption{\it Static Wilson loop with contour $\Gamma_0$.} 
\label{plwil}
\end{figure}

The quark-antiquark Green function is given by 
\begin{eqnarray}
G_{\beta_1\beta_2\alpha_1\alpha_2}(T) &{\buildrel {m_j\to\infty}\over \longrightarrow }& \,
\delta^3({\bf x}_1-{\bf y}_1) \delta^3({\bf x}_2-{\bf y}_2) (P_+)_{\beta_1\alpha_1} (P_-)_{\alpha_2\beta_2}
\nonumber \\
& & \times e^{-(m_1+m_2) T}\langle {\rm Tr}\, {\rm P}  e^{ig \oint_{\Gamma_0} dz_\mu A_\mu (z) }\rangle
\label{solgreen}
\end{eqnarray}
with $P_\pm\equiv  (1\pm \gamma_4)/2$. The integral in Eq. (\ref{solgreen}) extends over the circuit 
$\Gamma_0$ which is a closed rectangular path with spatial and temporal extension $r=\vert {\bf x}_1 -{\bf x_2}\vert$
and $T$ respectively, and has been formed by the combination of the path-ordered exponentials
along the horizontal (=time fixed) lines, coming from the Schwinger strings, and those along the 
vertical lines coming from the static propagators (see Fig. \ref{plwil}). The brackets in (\ref{solgreen}) denote 
the pure gauge vacuum expectation value. In Euclidean space
\begin{equation}
\langle f[A]\rangle  \equiv {1\over Z} \int {\cal D} A f [A] e^{-\int d^4 x L^E_{YM}}.
\label{expece}
\end{equation}

From Eq. (\ref{solgreen}) it is clear that the dynamics of the quark-antiquark interaction is contained in 
\begin{equation}
W(\Gamma_0) =  {\rm Tr\, P}  e^{\displaystyle i g \oint_{\Gamma_0} dz_\mu A_\mu (z) }.
\label{wilsstat}
\end{equation}
This is the fa\-mous ({\it static}) Wegner--Wilson loop (Wegner (1971) and Wil\-son (1974)). 
In the limit of infinite quark mass considered, the kinetic energies of the quarks drop out of the theory, 
the quark Hamiltonian becomes identical with the potential (see Ex. 4.1.1) while the full Hamiltonian contains also 
all types of gluonic excitations. According to the Feynman--Kac formula the limit $T\to \infty $ projects
out the lowest state i.e. the one with the ``glue'' in the ground state. This  has the role of the quark-antiquark 
potential for pure mesonic states. Now, comparing Eq. (\ref{solgreen}) with Eq. (\ref{inf}) and considering that 
the exponential factor $\exp{(- (m_1+m_2) T)}$ just accounts for the the fact that the energy of the 
quark-antiquark system includes the rest mass of the pair\footnote{Moreover $E(R)$ includes also 
self-energy effects which need to be subtracted when calculating the quark-antiquark potential.}, we obtain
\begin{equation}
V_0(r) \equiv E(r) = -\lim_{T\to \infty} {1\over T } \log \langle W(\Gamma_0) \rangle.
\label{potfond}
\end{equation}
The quark degrees of freedom have now completely disappeared and the expectation value in (\ref{potfond}) 
has to be evaluated in the pure Yang--Mills theory ((Wilson (1974), Brown and Weisberger (1979)).
Notice that the potential is given purely in terms of a gauge invariant quantity (the Wilson loop 
precisely). In this way we have reduced the calculation of the static potential to a well posed problem 
in field theory: to obtain the actual form of $V_0$ we need to calculate the QCD expectation value of the  static 
Wilson loop. 

We conclude this section pointing out that Eq. (\ref{potfond}) is rigorously true for static sources. 
For (realistic) heavy quarks with finite mass, the static potential, interpreted 
as the static limit of the potential appearing in the Schr\"odinger equation, could in principle 
not coincide exactly with Eq. (\ref{potfond}). Actually it does not. 
The reason is that in QCD quarks in the static limit can still change colour by emission 
of gluons. This introduces a new dynamical scale in the evaluation of the potential from 
Eq. (\ref{potfond}) of the order of the kinetic energy which is finite if the quarks  
have finite mass. In perturbative QCD this new scale is given by the difference between the 
singlet and the octet potential. Contributions of the same order of the kinetic energy are 
not of potential type and have to be explicitly subtracted out from Eq. (\ref{potfond}). 
In the next section we will give the leading effect of this subtraction on the static Wilson loop. 
For an extended analysis we refer to Brambilla et al. (1999). 

\vskip 1truecm
\leftline{\bf 4.1 Exercises}
\begin{itemize}
\item[4.1.1]{ Consider a particle of mass $m$ moving in a potential $V(x)$  in one space dimension. The propagator 
is given by $\displaystyle K(x^\prime, t; x, 0)= \langle x^\prime \vert \exp{(-i Ht)}\vert x\rangle$ with 
$H= \displaystyle{p^2\over 2 m} +V(x)$. Obtain the form of the propagator $K$ in the static limit 
$m\to \infty$ and compare it  with Eq. (\ref{soles}).} 
\item[4.1.2]{ Consider quenched QED. Show that the functional generator for the QED Lagrangian
with a source term added of the form $J_\mu(x) A_\mu(x) $ with $J_\mu(x) =e \delta_{\mu 4}
(\delta^3({\bf x} -{\bf x}_1)-\delta^3({\bf x} -{\bf x}_2)) $, coincides with the vacuum expectation 
value of the static Wilson loop in the limit of infinite interaction time.}
\end{itemize}

\subsection{The Wilson loop in perturbative QCD}
The Wegner--Wilson loop of contour $\Gamma$ is defined in QCD as 
\begin{equation}
W(\Gamma)\equiv {\rm Tr\, P}\,  e^{ig \displaystyle\oint_{\Gamma} d z_\mu A_\mu(z) }.
\label{wilver}
\end{equation}
Due to the presence of the colour trace, it is a manifestly gauge invariant object. The field $A_\mu$ can be taken 
in any representation of $SU(3)$.  When describing the quark-antiquark interaction, as in the present case,
$A_\mu\equiv A_\mu^a \lambda^a /2$.  The Wilson loop is called static when the integral is extended 
to a rectangular $\Gamma_0$ as in Fig. \ref{plwil}. Physically, in the static Wilson loop only the 
time component $A_4$ is relevant.

We know from the previous section that the vacuum expectation value of $W$ on the QCD measure gives  the 
static potential. However, we are in trouble when we step in to calculate it. Indeed, if we want to 
describe the long range quark-antiquark interaction,  we should be able to calculate the Wilson loop
in the region in which the running coupling constant is no longer small and we should be able to 
sum up all the relevant diagrams. Unfortunately, we have no methods at hand to sum up such contributions.  
Worse enough, also usual semiclassical approaches are not doomed to work in this case. 
We do not know of any dominant and confining configurations in the QCD measure and in the path integral that can make 
the work\footnote{Instantons do not confine directly, i.e. give a zero string tension. 
Monopoles arise in the Abelian projection or after a dual transformation, see Sec.6.}!
Hence, to obtain information on the behaviour of the Wilson loop in the nonperturbative region we have to resort 
either to strong coupling expansion and to lattice simulations, see Sec. 4.3 and Sec. 4.5, or to 
analytic models of the QCD vacuum, see Sec. 6. 
 
However, in the weak coupling region the Wilson loop can be calculated perturbatively.
Recently, the fully analytic calculation of the two-loop diagrams contributing to the static potential 
has been performed (Peter (1997), Schr\"oder (1999)). We refer to the original papers for the details of 
the calculation. However, due to the high interest, we report here the final result
($\alpha_{\rm s}$ is in the $\overline{MS}$ scheme):
\begin{eqnarray}
&& V_0(r) = -C_F{\alpha_V(r) \over r}\label{vr}\\
& & \alpha_V (r)  = \alpha_{\rm s}(r)
\left\{1+\left(a_1+ {\gamma_E \beta_0 \over 2}\right) {\alpha_{\rm s}(r) \over \pi}\right.
\nonumber\\
&&+\left.\left[\gamma_E\left(a_1\beta_0+ {\beta_1 \over 8}\right)+\left( {\pi^2 \over 12}+\gamma_E^2\right) 
{\beta_0^2 \over 4}+b_1\right] {\alpha_{\rm s}^2(r) \over \pi^2} \right \},
\nonumber
\end{eqnarray}
where $\gamma_E$ is  the Euler constant, $a_0=1$, 
$$
a_1 = \frac{31}{9}C_A - \frac{20}{9}T_FN_f
$$
and
\begin{eqnarray*}
a_2 &=& \Big(\frac{4343}{162}+4\pi^2-\frac{\pi^4}{4}+\frac{22}{3}\zeta_3\Big)C_A^2 
-\Big(\frac{1798}{81}+\frac{56}{3}\zeta_3\Big)C_AT_FN_f\nonumber \\ &&
-\Big(\frac{55}{3}-16\zeta_3\Big)C_FT_FN_f + \Big(\frac{20}{9}T_FN_f\Big)^2.  
\end{eqnarray*}
$C_F = 4/3$ and $C_A=3$ are the Casimir of the fundamental and of the adjoint representation 
respectively. Moreover in QCD we have $T_F = 1/2$.

From this result we learn that: the two-loop contribution is nearly as large as the one-loop term, 
both make the potential more attractive and eventually  the perturbative potential seems to be reliable 
up to a distance  $r\Lambda_{QCD} < 0.07$, which is considerably smaller than the average radius 
in quarkonia (see Fig. \ref{plpot}). For larger values a strong scale-dependence  remains and the perturbation 
series breaks down above 0.1 fm. Hence, even the pure perturbative calculation indicates the need of a different 
long range approach\footnote{In Pineda and Yndurain (1998) the two-loop static potential and the one-loop 
relativistic perturbative corrections to the potential were used in order to calculate the 
ground state energies of bottomonium and charmonium and thus to obtain a value for the bottom 
and charm masses. Nonperturbative effects were encoded in the local gluon condensate. In the next section, 
we will show that nonperturbative contributions are actually carried by non-local quantities, Gromes (1982), 
like the Wilson loop, which can be approximated by local condensates only if the involved physical scales 
enable a local expansion. This is indeed the case of the bottomonium ground state.}.

We mention that the next  perturbative correction to the static potential can be 
obtained only in an effective theory framework (pNRQCD, see Sec. 5). In that framework the leading log 
three-loop term has been very recently calculated in Brambilla et al. (1999). 
It amounts to a correction $C_A^3 \alpha_{\rm s}^4(r) / 12 \pi \log{ r \mu^\prime}$ to $\alpha_V$ in 
Eq. (\ref{vr}), $\mu^\prime$ being the scale of the matching. Notice that the potential comes to depend 
on the infrared scale $\mu^\prime$. This signals the appearance at three-loop of a nonpotential type of 
contribution to the static Wilson loop which has been subtracted out explicitly at $\mu^\prime$. 
 
In QED in the quenched approximation (i.e. neglecting light fermions) the (non-static) Wilson loop can be 
calculated analytically in a closed form. We sketch here the derivation. We have (in Euclidean space) 
\begin{equation}
\langle W(\Gamma)\rangle = { \displaystyle\int {\cal D} A \, e^{\displaystyle {1\over 2} 
\displaystyle\int d^4 x A_\mu\, M_{\mu \nu} A_\nu
+i e \oint dz_\mu A_\mu }\over \displaystyle\int {\cal D} A \,  e^{\displaystyle{1\over 2} 
\displaystyle\int d^4 x A_\mu\,  M_{\mu \nu} A_\nu}}
\label{wilqed}
\end{equation}
where $M_{\mu\nu} \equiv \delta_{\mu \nu } \partial^2 -\partial_\mu \partial_\nu $ has been obtained by 
integrating by parts the original QED action. The integral over the $A$ field in Eq. (\ref{wilqed}) 
is Gaussian and therefore can be performed provided that a gauge condition is imposed. However, since the 
Wilson loop is a gauge-invariant quantity, the choice of the gauge is immaterial.
We obtain
\begin{equation}
\langle W(\Gamma) \rangle =  \exp\left\{ {e^2\over 2} \oint_\Gamma dx_\mu \oint_\Gamma  
dx^\prime_{\nu} D_{\mu \nu}(x-x^\prime) \right\}
\label{solqed}
\end{equation}
where $D_{\mu \nu}(x-x^\prime) \equiv \langle A_\mu(x) A_\nu(x^\prime)\rangle$ is the photon propagator.
E.g. in Feynman gauge we have $D_{\mu \nu}(x) = - \displaystyle{\delta_{\mu\nu} \over 4\pi^2}{1\over x^2}$.  
On a rectangular Wilson loop Eq. (\ref{solqed}) becomes 
\begin{equation}
\langle W(\Gamma_0) \rangle = \exp\left\{{e^2\over 4 \pi r} T f(r,T)\right\}
\label{intwil}
\end{equation}
with $f(T,r) =\displaystyle{2\over \pi} \left[{\rm arctan}{T\over r} -{r\over 2 T}\log(1+{T^2\over r^2})\right] $, 
$f\to 1 $ for $T\to \infty$ getting back the Coulomb potential. 

\begin{figure}[htb]
\vskip -0.1truecm
\makebox[4.0truecm]{\phantom b}
\centerline{
\epsfxsize=5truecm\epsffile{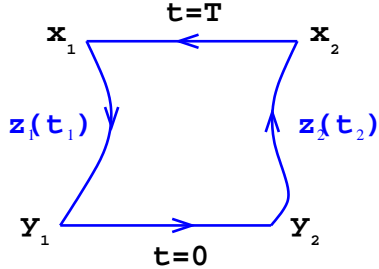}}
\vskip 0.1truecm
\caption{\it  Generalized Wilson loop with contour $\Gamma$.} 
\label{plwil2}
\end{figure}

In the  weak coupling region the vacuum expectation value of the (non-static) Wilson loop in QCD can be 
obtained by making a Gaussian approximation on the functional integral (i.e. by neglecting 
non-Abelian contributions).  In this case, since the contribution coming from 
the Schwinger strings  vanish in the limit $T \to \infty$ and taking advantage of 
the notation given in Fig. \ref{plwil2}, we have       
\begin{eqnarray}
\langle W(\Gamma) \rangle \! &=& \!
e^{\displaystyle{4\over 3} g^2 \displaystyle\oint_{\Gamma} dz^1_\mu \oint_{\Gamma} dz^2_\nu
D_{\mu\nu}(z_1 - z_2)}\nonumber \\
&{\buildrel  { T\to \infty} \over \longrightarrow}& \!
e^{\displaystyle{4\over 3} g^2 \displaystyle \int_{t_i}^{t_f} dt_1 \int_{t_i}^{t_f} dt_2 \dot{z}_\mu^1(t_1) 
\dot{z}_\nu^2(t_2) D_{\mu\nu}(z_1(t_1) - z_2(t_2))}\,   ,
\label{propp}
\end{eqnarray}
$D_{\mu\nu}$ being the gluon propagator.

\vskip 1truecm
\leftline{\bf 4.2. Exercises}
\begin{itemize}
\item [4.2.1]{ From Eqs. (\ref{potfond}) and (\ref{propp}) obtain the QCD one gluon exchange contribution 
to the static QCD  potential.}
\item [4.2.2]{ Calculate Eq. (\ref{intwil}) from Eq. (\ref{solqed}).}
\item [4.2.3]{ From the correspondent of Eq. (\ref{willexpp}) (Sec. 5) in Euclidean space, 
obtain the QCD $V_0$ and $V_{\rm VD}$ potentials using 
the weak coupling behaviour of the Wilson loop given in (\ref{propp})  with the gluon propagator 
first in the Coulomb gauge and then in the Feynman gauge.  Demonstrate that, due to the gauge 
invariance of the Wilson loop, the two expression coincide. [Hint: perform the change 
of variables $t=(t_1+t_2)/2; \tau=t_1-t_2$ in the integrals in  (\ref{propp}), expand ${\bf z}_j$ 
around $t$  and integrate over $\tau^{+\infty}_{-\infty}$.]}
\end{itemize}

\subsection{Lattice formulation, strong coupling expansion and area law}
Equation (\ref{potfond}) is particularly useful on the lattice where the dynamical 
variables are unitary matrices associated with the links. We do not want to give here an introduction 
to lattice QCD (for this we refer the reader to Rothe (1992) and Montvay and Munster (1994)). However, 
in order to illustrate some interesting results, we recall the basic definitions and concepts.

\begin{figure}[htb]
\makebox[4.0truecm]{\phantom b}
\epsfxsize=4truecm\epsffile{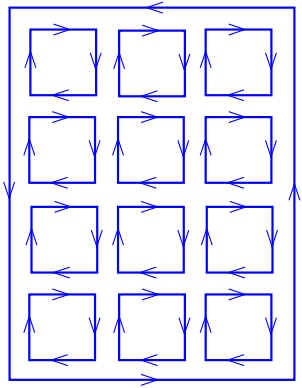}
\caption{\it Elementary plaquette on the lattice and tiling of the Wilson loop in strong coupling.} 
\label{plpla}
\end{figure}
 
Let us consider QCD in the pure gauge sector on a four-dimensional Euclidean discretized space-time, 
the  ``lattice'' of step $a$. Lattice sites are denoted by $n $ and lattice directions are denoted by $\mu, \nu$. 
We define the group element associated with a link, from the lattice site $n$ to $n+\hat{\mu}$ ($\hat{\mu} $  
being a unit vector along the axis $\mu$) as    
\begin{equation}
U_\mu(n)\equiv U(n,n+\hat{\mu})\simeq e^{ig a A_\mu(n+{\hat{\mu}\over 2 })}
\quad \quad  U^\dagger_\mu(n) \equiv U(n+\hat{\mu},n) .
\label{link}
\end{equation}
These are the  dynamical variables, the gluonic colour fields that relate the colour coordinate system 
at different space-time points. From (\ref{link}) we see that they are represented as a $3\times 3$ colour matrix 
that can be interpreted as the path-ordered exponential of the continuum colour fields $A_\mu(x)$. 
The  gauge transformation on the lattice, ${\cal V}(x_n)$, acts directly on the link elements
\begin{equation}
U_\mu(n) \to {\cal V}^\dagger(n+\hat{\mu}) U_\mu(n) {\cal V}(n) . 
\label{trasf}
\end{equation}
The Yang--Mills action is 
\begin{equation}
S= -{\beta\over 3} \sum_{\mu > \nu} \sum_n \, {\rm  Re}\,  {\rm Tr}\,  U_{\mu\nu}
\label{latac}
\end{equation}
with $\beta= 6/g^2$. The elementary  plaquette is the trace of the path-ordered 
product of links around a unit square (see Fig. \ref{plpla})
\begin{equation}
U_{\mu\nu}(n)\equiv U_\mu(n) U_\nu(n+\hat{\mu})U^\dagger_\mu(n+\hat{\nu})U^\dagger_\nu(n); 
\quad \quad U_P\equiv {1\over 3}  {\rm Re}\,  {\rm Tr} \, U_{\mu\nu} .
\label{defplaq} 
\end{equation}
In the continuum limit ($a\to 0$) expression (\ref{latac}) reduces to the usual Yang--Mills action\footnote{
Of course there exists an infinite number of lattice actions that have the same naive 
continuum limit, in particular when one considers also the fermion contribution. To be sure that 
the given lattice action reproduces really QCD for $a\to 0$, the lattice theory should exhibit  
a critical region in parameter space where the correlation length diverges. See also Sec. 4.5.}. 
The corresponding partition function is
\begin{equation}
Z= \int \prod_{n,\mu}  {\cal D}U e^{-{\beta} \sum_P U_P }
\label{part}
\end{equation}
where the integral is over the group manifold of colour $SU(3)$ for each link matrix $U$.

The Wilson loop is simply the trace of the product of the matrices $U(n,\mu) $ along the contour  $\Gamma$
which here is  a rectangle. Then 
\begin{equation}
\langle W(\Gamma) \rangle 
=\langle {\rm Tr} \, \prod_{l\in \Gamma} U(l,\mu_l) \rangle= {1\over Z} \int \prod_{n,\mu} DU(n,n+\hat{\mu}) 
{\rm Tr}\, \prod_{l\in \Gamma} U(l,\mu_l)  e^{-{\beta} \sum_P U_P}  .
\label{will}
\end{equation}

Of particular interest  is the strong coupling expansion on the lattice, which means to expand for large  
$g$ (small $\beta$). This bears a relation to the string picture that, as we explained,
characterizes the long range quark-antiquark interaction. 
In the strong coupling,  we can expand the exponential of the action in (\ref{will})
\begin{eqnarray}
& & \langle W(\Gamma) \rangle = {1\over Z} \int \prod_{n, \mu} DU(n,n+\hat{\mu}) 
{\rm Tr}\, \prod_{l\in \Gamma} U(l,\mu_l) \nonumber \\
& & \quad \times  \left[1-{\beta} \sum_P {\rm Tr}\, U_P +{1\over 2} {\beta }^2
\sum_P\sum_{P^\prime} {\rm Tr}\, U_P\,  {\rm Tr}\, U_{P^\prime} +\cdots \right]. 
\label{exp}
\end{eqnarray}
Since each plaquette in the expansion costs a factor $\beta$, the leading contribution in the limit 
$\beta \to 0$ is obtained by paving the inside of the Wilson loop  with the smallest number of elementary 
plaquettes yielding a non-vanishing value for the integral.
Using the orthogonality relations supplied in Exercise 4.3.1,  it is possible to show that 
the relevant configuration is the one presented in Fig. \ref{plpla} and that 
\begin{equation}
\langle W(\Gamma) \rangle \simeq \left({1\over g^2}\right)^{N_P} \,  , 
\label{area}
\end{equation}
$N_P$ being the minimal number of plaquettes required to cover the area enclosed by the path $\Gamma$
(for more details see Creutz (1983)).  This corresponds to the area law (Wilson (1974)) 
since the area enclosed by the path  $\Gamma$ is given by $A(\Gamma)=a^2 N_P$.
Furthermore, it is possible to demonstrate that the strong coupling expansion (\ref{exp}) has a finite 
radius of convergence. Hence, for $g^2$ large enough, the vacuum expectation value of the Wilson loop 
has the behaviour
\begin{equation}
\langle W(\Gamma) \rangle \simeq (g^2)^{-A(\Gamma)/a^2} = e^{-\displaystyle{r\, T \log g^2 \over a^2}}
\label{behav}
\end{equation}
where the last equality holds for a rectangular path $r \times T$. 
The behaviour of the Wilson loop given by Eq. (\ref{behav}) leads to a linear static potential with
\begin{equation} 
\sigma \simeq {\log g^2\over a^2}.
\label{sigimp}
\end{equation}
The layer of plaquettes giving this area contribution corresponds to a constant (chromo)electric field 
along the string connecting the quark and the antiquark. This suggests once again 
the relevance of a flux tube description of the nonperturbative interaction  and 
is at the origin of the formulation of the flux tube model of  Isgur et al. (1983). 

However, we have to keep in mind that the continuum limit of lattice QCD  is reached in the weak coupling 
limit, $\beta \to\infty$. Therefore it is impossible to extrapolate the strong coupling results directly 
to the continuum  physics. Still, we can argue that a rather coarse lattice with large lattice spacing 
should already give some indicative results for a theory {\it without} phase transition. To such a 
lattice corresponds a large bare coupling $g$ and hence the strong coupling result (\ref{behav}) 
should give the correct qualitative picture. Our philosophy is to take  the behaviour (\ref{behav})  
as a reliable suggestion to be used in the continuum physics. 

The leading order strong coupling  expansion of the Wilson loop in QED is identical to Eq. (\ref{behav}) 
and therefore produces  a linear  potential too. However, in the case of QED  on the lattice a phase transition  
is clearly seen (Kogut et al. (1981)) when going from strong coupling to weak coupling. 
In QCD the analytic proof that there cannot be such a phase transition together with 
the finite radius of convergence of the strong coupling expansion would be equivalent 
to {\it a proof of confinement}. Such a proof does not exist up to now. However, the numerical lattice 
simulations present no hint of such a transition in the intermediate coupling region. On the contrary, 
the strong coupled behaviour $g^2(a) \sim e^{\sigma a^2}$ continuously 
goes into the weak coupling $g^2(a) \sim 1 / \log{a^{-1}}$ as $a \to 0$.
Moreover, within the coupling regions accessible to present day computers, there are already 
overlaps  between lattice results and weak coupling expansion, the most impressive being the 
result of the alpha collaboration, see Capitani et al. (1998).

These results hint to the fact that QCD possesses  both the property of asymptotic freedom and colour 
confinement. In other words the Wilson loop in QCD displays a perimeter law in weak coupling 
and an area law in strong coupling.

\vskip 1truecm
\leftline{\bf 4.3  Exercises}
\begin{itemize}
\item[4.3.1]{ The orthogonality properties of the group integral in $SU(3)$  are given by
\begin{eqnarray*}
& & \int dU(n, n+\hat{\mu}) [U(n,n+\hat{\mu})]_{ij} =0 \nonumber \\
& & \int dU(n, n+\hat{\mu}) [U(n,n+\hat{\mu})]_{ij} [U^\dagger(n,n+\hat{\mu})]_{kl} = {1\over 3} \delta_{il} \delta_{jk}
 \nonumber \\
& & \int dU(n, n+\hat{\mu}) [U(n,n+\hat{\mu})]_{ij} [U(n,n+\hat{\mu})]_{kl} = 0 
\end{eqnarray*}
Using these relations justify the result (\ref{area}).}
\end{itemize}

\subsection{Area law as a criterium for confinement, \\ duality and the Wilson loop as an order parameter}
To see whether QCD shows confinement, one can study the energy of a system composed of a quark and an antiquark 
along the lines exposed in Sec. 4.1.  Then  Eq. (\ref{potfond}) tells us that it is the Wilson loop and 
its behaviour that determines the confinement property of the theory. In the previous section we have 
seen that in QCD in strong coupling expansion the Wilson loop in the fundamental representation 
obeys an area law behaviour  and this in turn via Eq. (\ref{potfond})  confirms 
the property of quark confinement. 

For very large loops $\log \langle W(\Gamma) \rangle$ generally exhibits these two types of behaviour: it decreases 
either as the perimeter or as the area of $\Gamma$. In the first case, expansive loops are allowed and quark and 
antiquark can be far apart from each other. In the second case quark and antiquark propagate as a bound state.
This is the Wilson criterion for confinement of electric charges (Wilson (1974))
\begin{eqnarray} 
\langle W(\Gamma) \rangle & & \sim e^{- K L(\Gamma)} \quad \quad~ {\rm no} \> {\rm confinement}\\
\langle W(\Gamma) \rangle & & \sim e^{- K^\prime A(\Gamma)} \quad \quad   {\rm confinement}
\label{order}
\end{eqnarray}
with $L(\Gamma)$ the perimeter of $\Gamma$,  $A(\Gamma)$ the minimal area enclosed by $\Gamma$, and $K$
and $K^\prime$  dimensionful 
constants. These are statements about the response of the pure gauge vacuum to external perturbations.

This criterion inspires physical pictures of the QCD vacuum (see Sec. 6). 
In particular it was suggested by 't Hooft (see e.g. 't Hooft (1994)) that a pure non-Abelian gauge 
theory could display quite a complicate pattern of vacuum phases: 1) The Higgs mode. Only colour magnetic 
charges are confined. 2) The Coulomb mode, featuring ordinary massless ``photons''  and no superconductivity or 
confinement. 3) The confinement mode or ``magnetic superconductor''.  Quarks (i.e. electric charges) are 
permanently confined. The properties of these phases are then expressed in terms of operators that create vortices of 
electric flux (Wilson loops $W_A(\Gamma)$, evaluated on the gauge fields $A_\mu$) and operators that create 
vortices of magnetic flux ('t Hooft operators $W_C(\Gamma)$, evaluated on the dual potentials $C_\mu$). 
Using the definition of  $W_C(\Gamma)$  via its commutation relation with $W_A(\Gamma)$,
't Hooft (1979)  showed that if $W_A(\Gamma)$ obeys an area law, then $W_C(\Gamma)$ necessarily satisfies a
perimeter law. The phase in which the 't Hooft operator obeys an area law  is the Higgs phase, while 
that one in which the Wilson loop obeys an area law is the confinement phase. 't Hooft result is a precise 
way of saying that the vacuum of a confining theory has the properties of a magnetic superconductor.  
The key ideas are the fact that a non-Abelian gauge theory can be viewed as an Abelian gauge 
theory enriched with Dirac magnetic monopoles and the concept of electric-magnetic duality. 
For QCD this implies that the vacuum behaves as a dual superconductor and confinement is explained 
through the monopole condensation, see Sec. 6. However, if in an Abelian theory  it is possible to introduce  dual 
field strengths and electric potentials $C^\mu$ dual to the ordinary (magnetic) potentials $A^\mu$ 
in a straightforward manner, in a non-Abelian theory  it is not possible to express the dual potentials 
explicitly in terms of ordinary potentials. The explicit form of the exact Yang--Mills Lagrangian  
as a function of the $C^a_\mu$ fields is unknown. The construction of the long distance limit of the Lagrangian 
is based on the fact that the dual potentials are weakly coupled since the dual Wilson 
loop obeys a perimeter law.  The quadratic part of the dual Lagrangian is thereby determined. 
The minimal  extension can be constructed under requisite of dual gauge invariance. For a concrete 
construction of a QCD dual Lagrangian see Baker et al. (1985)-(1995); Maedan et al. (1989) and Sec. 6.

These ideas have  originated a quite intensive research activity in the ``nonperturbative'' 
physics community in the last few years (in  QCD mainly on the lattice while in supersymmetric 
field theories very promising analytic results are obtained just now, see e.g. 
Alvarez-Gaum\'e et al. (1997)). We will come back to this point in Sec. 6.2. Here we refer the reader 
to the  historical papers of Nambu (1974), Mandelstam (1979) and  't Hooft (1982) and references therein.

\subsection{Lattice simulations and lattice results}
With  the action of Eq. (\ref{latac}), using periodic boundary conditions in space and time and 
taking a lattice of finite volume and finite spacing, the system has a finite 
number of degrees of freedom: the gluon fields on the link  and (if we add them) the quark fields at the lattice sites. 
The functional integral over the gauge fields is converted to a multiple 
integral with a positive definite integrand (in Euclidean space): Eq. (\ref{part}).  
Unlike in continuum perturbation theory the calculation of averaged values of gauge invariant 
observables is done without  any gauge-fixing. 

For a lattice of $L^4$ sites with colour group $SU(3)$ the integral 
in (\ref{part}) would be a $8\times 4 \times L^4$ dimensional integral:
the standard approach  is to use a Monte Carlo approximation to the integrand. 
Precisely, a stochastic estimate  of the integral is made  from a finite number of samples (the ``configurations'')
of equal weight. For more details see Rothe (1992).  Here we are interested only in explaining 
that in this way we have at our disposal a set of samples of the vacuum, then, it is possible 
to evaluate the average of fields over these samples and obtain a nonperturbative evaluation 
of  Green functions as well as of any field vacuum correlator\footnote{
Analytic continuation from Euclidean to Minkowski time of Green functions that are only available 
on a finite set of points is not possible. Therefore, on the lattice one can only determine masses 
and on-shell matrix elements in a straightforward way.  Real time processes like scattering, hadronic 
decays, are not directly accessible.}. 

In this section we present some lattice results on the calculation of the Wilson loop
expectation value, the static potential, the flux tube configuration and  the determination 
of the string tension $\sigma$. However, before, we comment briefly on the validity of the lattice results
and give few concepts to enable the ``reading'' of those results. 

The lattice is not the real world so that before extracting the physics we have to be sure that:  
1) the Green functions have  been extracted without contamination (the ground state mass should not be contaminated 
by pieces coming from the excited states);
2) the lattice size is big enough, we mean that the relevant physical distance should  fit in! (finite size error); 
3) the statistical errors of the calculation  is under control (statistical error); 
4) the lattice spacing is  small enough, since we have to approach the continuum limit (discretization error). 
On the other hand the lattice spacing supplies an explicit ultraviolet cut-off. 

The last condition is the most subtle one. Let us just sketch the idea. In order to extract the continuum limit from 
the lattice, it has to be shown that the results do not change if the lattice spacing is decreased further. 
However, the lattice spacing in physical units is not known directly. Actually it is measured. The lattice 
simulations are performed at a fixed value of $\beta$ (cf. Eq. (\ref{latac})). We recall that $\beta= 6/g^2$ and 
therefore a large $\beta$ corresponds to a small $g^2$. Ensembles at different values of $a$ are obtained 
by using different bare coupling constants $g$ in the action. However, the value of $a$ for a given value 
of $g$ is not known  {\it a priori} but has to be obtained calculating a dimensionful parameter and comparing it 
to an experiment. 

As an  example let us consider the string tension $\sigma$. Measured in lattice units it is only a 
function of the bare coupling: $\hat\sigma(g)$, the $\hat{}$ denoting a dimensionless quantity. 
In physical units, however, it has the dimension of a (mass)$^2$, so that the physical string tension 
is given by     
\begin{equation}
\sigma   = \lim_{a\to 0} {1\over a^2} \hat{\sigma}(g(a)).
\label{sigfis}
\end{equation}
It approaches a finite limit for $a\to 0$, if $g$ is tuned with $a$ in an appropriate way.
By requiring that in this limit 
\begin{equation}
a{\partial \over \partial a}\sigma =0
\label{eqdif}
\end{equation}
we obtain 
\begin{eqnarray}
\sqrt{\sigma}&=& c_\sigma \Lambda_L \nonumber\\
\Lambda_L &=& a^{-1} e^{-1\over 2 \beta_0 g^2} (\beta_0 g^2)^{-\beta_1\over 2 \beta_0^2} (1+ O(g^2))
\label{el}
\end{eqnarray}
where we have used the usual lattice convention for the beta function: $\beta(g)$ $=$ $-\beta_0 g^3$ 
$-\beta_1 g^5$  $+ O(g^7)$ (which differs for a factor $1/(4\pi)^2$ from Eq. (\ref{betaqcd})) 
and $\Lambda_L$ is independent of $a$ and fixes  the mass scale of the theory\footnote{The appearance 
of a scale  as $\Lambda_L$ is well known from perturbative continuum QCD, where the necessity 
of renormalizing the theory also requires the introduction of a scale. Usually the relation  between
$\Lambda_{QCD}$ in different regularization schemes and $\Lambda_L$ is known in  perturbation theory 
in the bare coupling. See however S. Capitani et al. (1998).}. 
This equation tells us again that a perturbative calculation 
of $\sigma$ is a priori impossible due to the non-analytic behaviour in $g$ of Eq. (\ref{el}).
It is possible to show that all  dimensionful  physical quantities can be expressed in the form
\begin{equation}
\theta_{phys}= \lim_{a\to 0} {1\over a^{d_\theta}}\hat{\theta}(g(a),a)=c_{\theta} (\Lambda_L)^{d_\theta}
\label{thet}
\end{equation}
$d_\theta$ being the naive dimension. Then, for small lattice spacing  we can determine  the universal
function
$g=g(a) $ by fixing the l.h.s. of (\ref{el}) at the physical value of the string tension. 
This gives $g$ as a function of $a \sqrt{\sigma}$
\begin{equation}
g^2(a)= - {1\over 2 \beta_0 \log(a \Lambda_L)}
\label{behavg}
\end{equation} 
(where $\Lambda_L$ is now determined by the physical condition imposed)
which ensures the finiteness of any observable and allows us to convert them  to physical 
units\footnote{A corresponding statement is expected to hold if the action depends 
on several parameters, e.g. coupling constant and quark masses.}.

The continuum limit has to be taken at a constant physical volume $L^4$. As $a$  is
decreased the number of lattice points $\hat{L}={L/a}$ increases. Therefore, a finite 
computer limits the lattice spacing to $a \geq a_{\rm min} > 0$.  In practice one is looking 
for scaling of the results within this window  or at least checking that the results follow 
the expected leading order $a$ dependence in order to safely extrapolate them to $a=0$.
A typical value for lattice simulations of QCD is $\beta \simeq 6$ and hence the bare coupling 
constant is $\alpha_{\rm s}= g^2/ (4 \pi) \simeq 0.08$.  In the early years of lattice QCD  it was widely assumed 
that at least for $\beta \simeq 6$, two-loop perturbation theory can be applied to $a(g)$. Therefore, after having 
determined $a$ at a coupling $g$, the $\Lambda$ parameter was extracted and $a(g\prime)$ for 
$g\prime \neq g$ computed via perturbation theory. Nowadays, the lattice spacing is determined 
separately for each simulation point $\beta$ by inputting one experimental value. In doing so,
$a(g)$ is obtained as a function of the bare lattice $g$. One finds big deviations from perturbation 
theory (asymptotic scaling) that are related  to the importance of the so called tadpole diagrams
in lattice perturbation theory (see e.g. Michael (1997), Davies (1997)
and Lepage et al. (1993) for these developments)\footnote{One might ask if it was  
nonetheless possible to invert  the $a(g)$  relation to obtain an $\alpha(q)$ and run the $q$ to high momenta 
subsequently. For this purpose effective couplings other than $g$ have been suggested and determined 
which show an improved asymptotic scaling behaviour.}.

The actual methods of simulating lattice QCD and extracting physical information have reached 
quite a high level of sophistication and we refer the interested reader to the reviews (e.g. Rothe (1992), Davies 
(1997), Montvay and Munster (1994)). The above discussion should be sufficient to make clear that: 1) lattice simulations 
are performed at a fixed value of $\beta$, 2) the value of $\beta$ is connected with the lattice spacing $a$,  3) 
results relevant for continuum physics are effectively independent of $a$ (scaling) 4) the extraction of the physical 
results requires to fix $a$ and in the quenched approximation it may be dependent on the experimental quantity 
chosen to fix it. It is, therefore, important to know what quantity was chosen when looking at the lattice data. 

\begin{figure}[htb]
\makebox[1.0truecm]{\phantom b}
\centerline{
\epsfxsize=8truecm
\epsffile{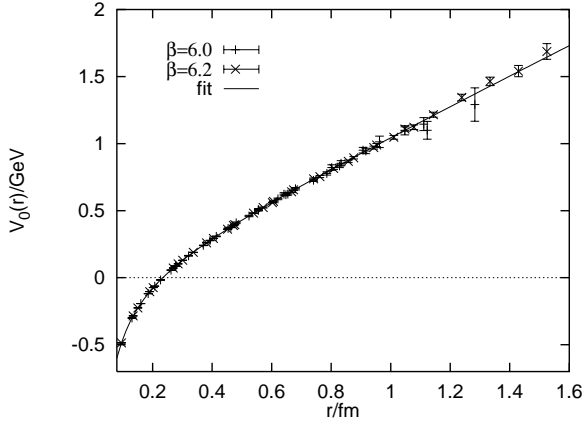}}
\vskip 0.1truecm
\caption{\it  Static potential at $\beta= 6.0$ and $\beta=6.2$ in the quenched approximation. 
The solid line is a fit of the form of Eq. (\ref{paramm}). Bali et al. (1997).} 
\label{plpotlat}
\end{figure}

Now let us come to some results.  First, we discuss the lattice measure of $\sigma$.
The lattice counterpart of Eq. (\ref{potfond}) is 
\begin{equation}
\hat{V}(\hat{r})= -\lim_{\hat{T}\to \infty} {1\over \hat{T}} \log W(\hat{r},\hat{T})
\label{latwilexp}
\end{equation}
where $ W(\hat{r},\hat{T})$ denotes the expectation value of a Wilson loop with spatial and temporal 
extension $\hat{r}$  and $\hat{T}$ respectively. Assuming that 
$ W(\hat{r},\hat{T})= \exp\{ \hat{\sigma} \hat{r} \hat{T} -\hat{\delta} (\hat{r}+\hat{T}) +\hat{\gamma}\} $,
the famous Creutz ratio 
\begin{equation}
\chi(\hat{r},\hat{T}) =-\log \left({W(\hat{r},\hat{T}) W({\hat{r}} - 1,{\hat{T}}-1)\over W(\hat{r},{\hat{T}}-1)
W({\hat{r}}-1,\hat{T})}\right)
\end{equation}
coincides with the string tension, since the parameters $\hat{\delta}$ and $\hat{\gamma}$ drop out.
The original  calculation (Creutz et al. (1982))
was performed on a $6^4$ lattice and the asymptotic scaling seemed to show up for values of $\beta$ slightly 
below 6.0. This measurement of a string tension different from zero from the strong coupling region to the asymptotic 
scale region, without any indication of a phase transition in the intermediate region, constituted the first 
{\it evidence} of quark confinement in QCD. 

In Fig. \ref{plpotlat} we show the most recent lattice measurement  
of the  quenched static potential from Eq. (\ref{latwilexp}) on a  hypercubic lattice $V= 16^4$ 
at $\beta=6.0$ and $V=32^4$ at $\beta=6.2$. These values correspond to  inverse lattice 
spacings $a^{-1}\simeq 2.1$ GeV and $a^{-1}\simeq 2.9$ GeV respectively.  
The scale is adjusted to optimally reproduce the bottomonium level splittings
(Bali et al. (1997)). The fit curve corresponds to the parameterization
\begin{equation}
V_0(r)= -{e\over r} +\sigma r + {f\over r^2}
\label{paramm}
\end{equation}
which clearly confirms  the Cornell potential (\ref{cornell}) (the $1/r^2$ correction, that accounts 
for the running of the coupling, is not meant to be physical but has been introduced to effectively 
parameterize the data  within the given range of $r$). The parameters take the value: $e= 0.321$ 
and $\sigma= (468$ MeV$)^2$. The coefficient of the Coulomb term is quite far from 
the effective value for $\alpha_{\rm s}$ coming from the potential models\footnote{Part of this discrepancy 
can be traced back to the quenched approximation.}.  Notice that the lattice potential becomes clearly 
linear around $0.2$ fm. Similar lattice measurements exist for the unquenched static potential (see e.g. Bali (1998)):
the string is found to break down around a quark-antiquark distance of about 1.2 fm. 

The static potential in Fig. \ref{plpotlat} has been extracted  as the ground state energy 
of the quark-antiquark configuration (cf. Eq. (\ref{fey})), 
which in turn corresponds to the lowest energy configuration of the ``glue'' between the quarks. 
Yet, also the excited gluonic modes have been measured on the lattice and the corresponding potentials 
have been adiabatically extracted. This should corresponds to the potential of heavy hybrids. 
For further details of this confirmation of the excited structure of the flux tube we refer 
the reader to Michael (1997), Morningstar et al. (1998).

\begin{figure}[htb]
\vskip -0.2truecm
\makebox[1.0truecm]{\phantom b}
\centerline{
\epsfxsize=10truecm
\epsffile{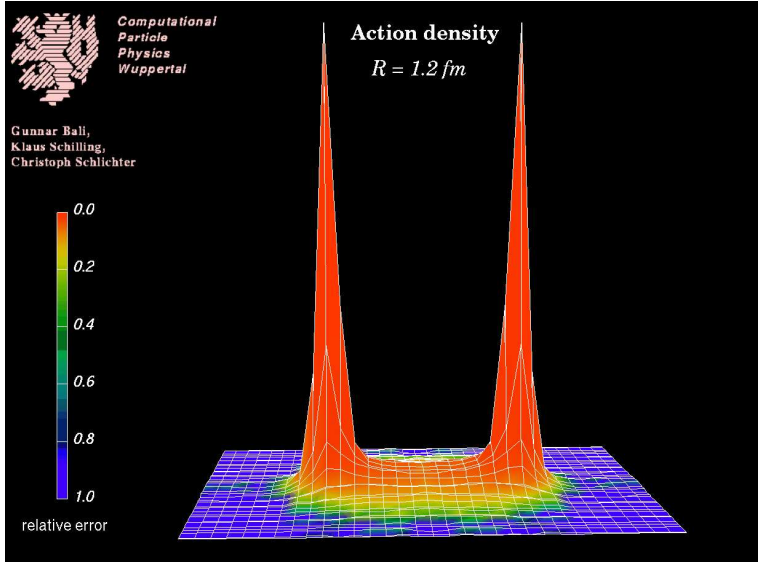}}
\vskip 0.5truecm
\caption{\it Action density distribution between two static quarks in $SU(2)$ measured 
on a lattice $V=32^4$ at $\beta=2.5$.  The physical quark-antiquark 
distance is 1.2 fm. Bali et al. (1995).}
\label{plac}
\end{figure}

Lattice studies have been  undertaken  to probe the energy momentum tensor of the colour fields.
The probe used is the gauge-invariant  insertion of a plaquette in the presence of a  static Wilson loop.  
Depending on the Lorentz orientation of the plaquette, this corresponds to the average 
of a (chromo)electric or a  (chromo)magnetic field in the presence 
of quark sources. Lattice sum rules can be used to normalize these distributions and to relate 
them to the $\beta$ function. For separations $r> 0.7$ fm, a string-like spatial distribution is found 
with a transverse rms  width increasing very slowly with $r$ and reaching a rather constant value between 1 and 2 fm.  
The physical value for this constant ranges between 0.5 and 0.75 fm. The averages of the squared components 
of the colour fields (which are gauge invariant quantities) are found to be roughly equal 
(i.e. $\langle E^2_i\rangle \sim \langle B^2_j\rangle$ in the presence of the Wilson loop, for $i,j=1,2,3$). 
This implies that the energy density is much smaller than the action density (in Euclidean space). 
(See Sec. 6.3 for a discussion.) The result for the action density is  presented in Fig. \ref{plac} 
and is quite impressive! The formation of the interquark flux tube is evident. The interpretation 
of this phenomenon is the following. When the distance between the quarks 
becomes larger than some critical value (connected to $\Lambda_{QCD}$)  the branching of the gluons,
due to the non-Abelian nature of QCD, becomes so intensive that it makes no sense to speak about individual 
gluons. A coherent effect develops with the subsequent formation of the flux tube.
It is conjectured that a specific organization of the QCD vacuum makes this kind of configuration 
energetically favorable (see Sec. 6).

However, lattice QCD seems more suitable to ask ``what'' and not ``why''. Here, we have shown  some 
model independent results on the quark interaction that provide evidence of quark confinement but do not explain 
{\it why} quarks are confined. Eventually, to get some insight in the quark confinement mechanism, it is 
necessary to build up and use some analytic models. The role of the lattice measurements will then  be  
to validate these models. The combination of analytic and lattice techniques will give us 
some information on the nature of the nonperturbative quark interaction.

Finally, we mention that there are a lot of measurements of hadron masses performed in lattice QCD
as well as in lattice nonrelativistic QCD (NRQCD). However, since in these lectures we are  more interested in 
the mechanism of confinement as well as in developing analytic approaches to it, we refer the reader 
to the literature (see e.g. Davies (1997)).

\section{The Heavy Quark Interaction}
In this part of the lectures we summarize some nonperturbative analytic results on the heavy quark 
interaction. With nonperturbative we mean the fact that they do not rely necessarily  on a  perturbative 
expansion in $g$. An analytic result of this type was already obtained for the static potential in Sec. 4 
with Eq. (\ref{potfond}). Here, we want to establish a systematic procedure to obtain relativistic corrections. 
In order to take full advantage from lattice calculations, our analytic approach will be manifestly gauge-invariant. 
In the next sections we present some model independent results: 1) the effective theories  of QCD 
that enable the  performing of a systematic expansion of the heavy quarks dynamics in some small parameters 
which are nonperturbative and 2) the exact and physically transparent expression for the quark-antiquark interaction 
at order $1/m^2$.
   
However, at the very moment we want to calculate the quark dynamics we have to resort to lattice evaluations 
or to models of the QCD vacuum (see Sec. 6). Nevertheless, the present approach allows us to gain 
something with respect to the standard lattice formulation. In fact, here  the lattice simulations come in 
at an intermediate step for the evaluation of some definite expectation values of fields inserted in 
the static Wilson loop. These can be directly tested with the analytic (model dependent) results.  
From this comparison, in a process of model validation, we gain a deeper understanding of the confinement physics.

A remark: we have performed all the  calculations of the last section in Euclidean space 
since lattice simulations (as all numerical evaluations) are done in Euclidean space.
We no longer  have this restriction  in the next sections where we come back to the Minkowski space.   
We use the following notation for the vacuum expectation value (to be compared with Eq. (\ref{expece}))
\begin{equation}
\langle f[A]\rangle \equiv {1\over Z} \int {\cal D} A f [A] e^{i \int d^4 x L_{YM}}.
\label{expecm}
\end{equation}
In most of the cases it will be  straightforward  to switch from Euclidean to Minkowski space by simply 
using the relations (\ref{euclidep}).

\subsection{Effective theories for heavy quarks}
We have seen that the physics of heavy quark bound states is complicated by the interplay  of 
different characteristic scales like the heavy quark mass $m$, the momentum of the bound state $mv$, 
the energy of the bound state $mv^2$ (and in principle also $\Lambda_{QCD}$), being $v$ the heavy 
quark velocity.  These scales can be disentangled using effective theories of QCD. 
From the technical point of view, this simplifies considerably the calculation. 
From the conceptual point of view, this enables us to factorize the 
part of the interaction that we know (and we are able to calculate in perturbation theory) 
from the low energy part which is dominated by nonperturbative physics.
 
Nonrelativistic QCD (NRQCD) is an effective theory equivalent to QCD and constructed 
integrating out the high energy ($E>m$) degrees of freedom, thus making explicit the mass parameter.
NRQCD has been extensively discussed at this school by Peter Lepage.  
We recall only few  points useful  to prepare the developments of the next sections. The interested 
reader is referred to Lepage (1996) and Thacker et al. (1991). 

NRQCD was devised to be applied to lattice simulations, however, here we are  interested in the definition 
of NRQCD in the continuum. In Sec. 4 we have considered the static limit of an infinitely massive  quark. 
In that case the Dirac equation for the quark propagator in an external field is exactly solvable. 
Now, we want to calculate the subsequent corrections showing up when the quark mass is large but 
finite. In order to  disentangle the dynamical scales it is convenient  to use  the heavy quark velocity,  
$v_Q\equiv v$, as an expansion parameter. Notice that, even if this seems to be  analogous 
to what happens in QED where, e.g. in positronium, the expansion parameter is $v_e \simeq \alpha$, 
the quark velocity is also sensitive to the nonperturbative quark interaction and turns out to be a function 
of both $\alpha_{\rm s}$ and $\Lambda_{QCD}$ (or $\sigma$).
   
Let us consider Eq. (\ref{direq}) (now in Minkowski space) or equivalently the Lagrangian 
\begin{equation}
L= \bar{\psi} (i \gamma^\mu D_\mu -m) \psi\,  .
\label{lcri}
\end{equation}
Applying a Foldy--Wouthuysen transformation and expanding in the inverse of the quark mass $m$, we 
obtain\footnote{We neglect here operators involving more than 2 fermions. These are irrelevant to our 
discussion here, see Brambilla et al. (1998).} 
up to order $1/m^2$ 
\begin{eqnarray}
L &=&  \psi^{\dag} \left( m + iD_0 +\frac {{\bf D}^2} {2m} 
+ c_1(m/\mu) \frac {({\bf D}^2)^2} {8 m^3} + c_2(m/\mu) \frac {g} {8 m^2}
({\bf D}\cdot {\bf E} - {\bf E}\cdot {\bf D}) \right.\nonumber\\
&& \left. + i  c_3(m/\mu) \frac {g} {8m^2} {\bfsigma} \cdot ( {\bf D} \times {\bf E} - 
{\bf E} \times {\bf D}) + c_4(m/\mu) \frac {g} {2m} {\bfsigma}\cdot {\bf B} + \dots\right)\psi \nonumber \\
& & + \hbox{antiquark terms} 
\label{nrqcd}
\end{eqnarray}
In expanding the QCD Lagrangian in the inverse of the mass we have lost explicit renormalizability. 
The NRQCD Lagrangian (\ref{nrqcd}) must be regularized. $\mu$ is an 
ultraviolet cut-off which restricts the momenta to the region $p \sim m v < \mu < m$.
The effect of the excluded momenta, e.g. in gluon loops, is factorized  in the (matching) coefficients 
which multiply the nonrelativistic operators. In Eq. (\ref{nrqcd}) we call them   
$c_i(m,\mu)$. Such coefficients are straightforwardly calculated in perturbation 
theory  by imposing that  scattering amplitudes evaluated with  (\ref{nrqcd}) are equal to the same amplitudes 
evaluated in QCD order by order in $\alpha_{\rm s}$ and in $1/m$ (see Manohar (1997)).
This procedure is called {\it matching}. The coefficients $c_i$ are normalized in such a way 
to be one (or zero) at tree-level.   

The terms in the Lagrangian can be ordered in powers of the squared velocity of
the heavy quark using the power counting rules for momentum and kinetic energy of 
Thacker et al. (1991):  ${\bf D} \sim  p \sim m v$,  $K \sim m v^2$. 
From the lowest order field equation 
$$
\left( i\partial_0 - g A_0 - \frac {{\bf D}^2} {2 m}\right) \psi = 0
$$
we get 
\begin{eqnarray*}
g A _0 \sim \partial_0 \sim K  = m v^2\\
g {\bf E} = [ D_0, {\bf D}] \sim pK = m^2v^3\\
-ig\epsilon_{ijk}B^{k} = [D_i, D_j] \sim K^2 = m^2v^4.
\end{eqnarray*}
Following these power-counting rules the number of operators to be included in $L$ can be truncated 
at a fixed order in $v^2$ depending on the precision we require on the calculation of the energy.

This power counting is not exact. It accounts only for the leading binding contributions. 
The reason is that two scales, the soft one $\sim mv$ and the ultrasoft $\sim m v^2$, are still 
mixed up in the NRQCD Lagrangian. An exact power counting can be achieved by integrating 
out from NRQCD the soft scale with the same procedure as the mass scale was integrated out from QCD 
in order to get NRQCD. In this way one obtains a further effective theory, called potential 
nonrelativistic QCD (pNRQCD) (Pineda and Soto (1998)) where only dynamical ultrasoft degrees 
of freedom are present. These are the heavy quark bound states (explicitly projected 
in colour singlet and octet states) and gluons propagating at the ultrasoft scale. 
Since nonrelativistic potentials get contributions only from the soft scale, 
in the pNRQCD Lagrangian potential and nonpotential contributions are explicitly disentangled.  
More precisely, the pNRQCD Lagrangian density with leading nonpotential corrections 
is given by (Brambilla et al. (1999)):
\begin{eqnarray}
L  &=& {\rm Tr} \left\{ S^\dagger \left( i\partial_0 - {{\bf p}^2\over m} 
+ \sum_n {V^{(n)}_s({\bf r},{\bf p},\bfsigma; \mu^\prime)\over m^n} \right) S \right\} 
\nonumber\\
& & + {\rm Tr} \left\{  O^\dagger \left( iD_0 - {{\bf p}^2\over m} 
+ \sum_n {V^{(n)}_o({\bf r},{\bf p},\bfsigma; \mu^\prime)\over m^n} \right) O \right\}
\nonumber\\
& & + g V_A ({\bf r};\mu^\prime) {\rm Tr} \left\{  O^\dagger {\bf r} \cdot {\bf E} \,S
+ S^\dagger {\bf r} \cdot {\bf E} \,O \right\} 
\nonumber\\
& & + g {V_B ({\bf r};\mu^\prime) \over 2} {\rm Tr} \left\{  O^\dagger {\bf x} \cdot {\bf E} \,O
+ O^\dagger O {\bf r} \cdot {\bf E}  \right\} 
\label{pnrqcd0}
\end{eqnarray}
where ${\bf r} \equiv {\bf x}_1-{\bf x}_2$ and  ${\bf X} \equiv ({\bf x}_1+{\bf x}_2)/2$,   
$S = S({\bf r},{\bf X},t)$ and $O = O({\bf r},{\bf X},t)$ are the singlet and octet 
wave function respectively. All the gauge fields in Eq. (\ref {pnrqcd0}) are evaluated 
in ${\bf X}$ and $t$. In particular ${\bf E} \equiv {\bf E}({\bf X},t)$ and 
$iD_0 O \equiv i \partial_0 O - g [A_0({\bf X},t),O]$. The matching coefficients 
$V_A$ and $V_B$ are normalized in such a way that at the leading perturbative 
order they are one. $\mu^\prime$ is the ultrasoft  cut-off corresponding to the 
regularization of the Lagrangian (\ref{pnrqcd0}): $m v$ $> \mu^\prime >$ $m v^2$.
At the leading order in the soft scale the equations of motion associated with the pNRQCD Lagrangian 
(\ref{pnrqcd0}) are two decoupled nonrelativistic Schr\"odinger equations describing the propagation 
of a singlet and an octet bound state defined by the potentials $\sum V_s^{n}/m^n$ and 
$\sum V_o^{n}/m^n$ respectively. Next-to-leading nonpotential corrections 
involve the coupling of octet and singlet states via ultrasoft chromoelectric fields. 
We point out  that: 1) The potentials have now the status of matching coefficients (in the matching 
from NRQCD to pNRQCD) and depend in general on the scale $\mu^\prime$ (see Sec. 4.2).  
They can be calculated comparing Wilson loop functions in NRQCD and singlet/octet
propagators in pNRQCD. The novel feature is that the matching takes place now in the low energy 
region and thus it can be even nonperturbative. In this case the potentials are given 
by expressions to be evaluated on the lattice. Notice that the potentials contain in their 
definition also the NRQCD coefficients $c_i(m/\mu)$. 2) the pNRQCD Lagrangian contains both potential 
and nonpotential contributions. In some kinematic regions they coincide with 
the Voloshin (1979) and Leutwyler (1981) corrections. 3) Only the ultrasoft scale is left. 
Hence each term in (\ref{pnrqcd0}) has a definite power counting in $v$! 
In particular: $1/r \sim p \sim m v$, $V^{(n)}_{s,o} \sim m v^2$ and $F_{\mu\nu} \sim m^2 v^4$.  

In the lattice NRQCD approach the NRQCD Lagrangian of Eq. (\ref{nrqcd}) is discretized on the lattice and 
the heavy  hadron masses are obtained by performing lattice simulations. These are considerably less time 
consuming than the traditional QCD lattice simulations because the cut-off scale $\mu$ (smaller 
than $m$) makes possible to use coarse lattices. Since in the following we will not deal explicitly 
with nonpotential contributions we will not perform explicitly the matching from NRQCD to pNRQCD.  
As far as ultrasoft degrees of freedom are not considered, this corresponds only to a choice of 
language. Actually, getting the heavy quark potential from the NRQCD Lagrangian, in any way one is doing it, 
if done properly, is nothing else than performing the leading order matching with the pNRQCD Lagrangian. 
In particular we will show how to obtain the spin and velocity dependent singlet potentials 
up to  order $1/m^2$. A possible method is to calculate these as corrections to the static propagator 
from the Lagrangian (\ref{nrqcd}) (see Eichten et al. (1981), Tafelmayer (1986)).
We will use a path integral formalism (Peskin (1983)). This approach has the advantage that the result, 
expressed in terms of deformed Wilson loops, is suitable to be used also in QCD vacuum models (see Sec. 6). 

One remark at the end. The matching coefficients $c_i$ contain all the fermionic loop 
contributions ~at the hard ~scale. If the matching ~between NRQCD and pNRQCD is nonperturbative, 
then the  quark loop contributions at the low energy scale are contained in the functional
measure of the object to be evaluated on the lattice (and it is indeed a hard task even 
for the lattice up to now). In the following section for simplicity we work in the quenched 
approximation.

\subsection{The QCD spin-dependent and velocity-dependent potentials}
Let us consider Eq. (\ref{nrqcd}), taking the matching coefficients at tree level\footnote{
The matching coefficients can be simply included in the calculation  (see Chen et al. (1995),
Bali et al. (1997)). They are needed for example to obtain the one-loop perturbative behaviour of the 
potentials, see  Brambilla et al. (1998)).}. From Eq. (\ref{nrqcd}) it follows that the heavy quark 
(of mass $m_j$) propagator $K_j$ in external field  (that now is reduced to  a Pauli quark propagator, 
i.e. a $2\times 2$ matrix in the spin indices) satisfies the Schr\"odinger equation
\begin{eqnarray}
& &i \frac{\partial}{\partial x^0} K_j(x,y;A) =  H_{{\rm FW}} K_j(x,y;A)   
\nonumber\\
& & \equiv \left[ m_j +\frac{1}{2m_j} ({\bf p}_j - g{\bf A})^2 
 -  \frac{1}{8m_j^3} ({\bf p}_j - g{\bf A})^4 - \frac{g}{m_j}
{\bf S}_j \cdot {\bf B} + gA^0 \right.  
\label{simpeq}\\
& & - \left. \frac{g}{8m_j^2} (\partial_i E^i - ig [A^i,E^i]) +  \frac{g}{4m_j^2} \varepsilon^{ihk} S_j^k
\{(p_j -gA)^i,E^h\} \right] K_j(x,y;A) 
\nonumber
\end{eqnarray}
with the Cauchy condition
\begin{equation}
K_j(x,y;A) |_{x^0=y^0} = \delta^3({\bf x}-{\bf y})
\label{incon}
\end{equation}
where $\varepsilon^{ihk}$ is the three-dimensional Ricci symbol. 
By standard techniques the solution of Eq. (\ref{simpeq}) (see Sakurai (1985) and Exercise 5.2.1)  
with the initial condition (\ref{incon}), can be expressed as a path integral in the phase space
\begin{equation}
K_j(x,y;A)= \int_{{\bf z}_j(y^0)={\bf y}}^{{\bf z}_j(x^0)={\bf x}}
{\cal D} [{\bf z}_j, {\bf p}_j] \, {\rm T} 
\exp \left\{ i \int_{y^0}^{x^0} dt \, [{\bf p}_j \cdot \dot{{\bf z}}_j - H_{{\rm FW}}] \right\}. 
\label{pathkappa}
\end{equation}
Here, the time-ordering prescription T acts both on spin and gauge matrices. The trajectory of the quark $j$ 
in coordinate space is denoted by ${\bf z}_j = {\bf z}_j(t)$, the trajectory in momentum 
space by ${\bf p}_j = {\bf p}_j(t)$ and the spin by  ${\bf S}_j$ (See Fig. \ref{plwil2}).
Standard path integral manipulations on Eq. (\ref{pathkappa}) give 
\begin{eqnarray}
K_j(x,y;A) \!\! &=& \!\! \int_{{\bf z}_j(y^0)={\bf y}}^{{\bf z}_j(x^0)={\bf x}}\!\!\!\!
{\cal D} [{\bf z}_j,{\bf p}_j] \, {\rm T} 
\exp \left\{ i \int_{y^0}^{x^0} \!\! dt \, \left[ {\bf p}_j \cdot \dot{{\bf z}}_j - m_j -
\frac{{\bf p}_j^2}{2m_j} + \frac{{\bf p}_j^4}{8m_j^3} 
\right. \right.
\nonumber\\
& &\qquad  - gA^0 + \frac{g}{m_j} {\bf S}_j \cdot {\bf B} + 
\frac{g}{2m_j^2} {\bf S}_j \cdot ({\bf p}_j \times {\bf E}) -
\frac{g}{m_j} {\bf S}_j \cdot (\dot{{\bf z}}_j \times {\bf E}) 
\nonumber\\
& & \qquad \left. \left.  + g \dot{{\bf z}}_j \cdot {\bf A} +
\frac{g}{8m_j^2} (\partial_i E^i -ig[A^i,E^i]) \right] \right\} .
\label{stran}
\end{eqnarray}

Inserting Eq. (\ref{stran}) into expression (\ref{utile}) of the quark-antiquark Green function and   
taking $x_1^0=x_2^0=t_{\rm f}$, $y_1^0=y_2^0=t_{\rm i}$ with $T\equiv t_{\rm f}-t_{\rm i} >0$, 
one obtains the two-particle Pauli-type propagator $K$ in the form of a path integral 
on the world lines of the two quarks
\begin{eqnarray}
K({\bf x}_1, {\bf x}_2, {\bf y}_1, {\bf y}_2;T)=
\int_{{\bf z}_1(t_{\rm i})={\bf y}_1}^{{\bf z}_1(t_{\rm f})={\bf x}_1}
    {\cal D}  [{\bf z}_1, {\bf p}_1]
\int_{{\bf z}_2(t_{\rm i})={\bf y}_2}^{{\bf z}_2(t_{\rm f})={\bf x}_2}
    {\cal D}   [{\bf z}_2, {\bf p}_2]
\nonumber\\
\times \exp\left\{i\int_{t_{\rm i}}^{t_{\rm f}} dt\, \sum_{j=1}^2
\left[{\bf p}_j \cdot \dot{{\bf z}}_j -m_j-
\frac{{\bf p}^2_j}{2m_j}+\frac{{\bf p}^4_j}{8m_j^3}\right] \right\}
\nonumber\\
\times \left\langle \frac{1}{3}
{\rm Tr \, T_s \, P} \exp\left\{ig\oint_{\Gamma} dz^{\mu} \,
A_{\mu}(z) +\sum_{j=1}^2\frac{ig}{m_j} \int_{{\Gamma}_j} dz^{\mu}
\right. \right.
\nonumber\\
\left. \left. \times \left(S_j^l \hat{F}_{l{\mu}}(z) -\frac{1}{2m_j} S_j^l\varepsilon^{lkr}p_j^k F_{{\mu}r}(z)-
\frac{1}{8m_j} D^{\nu}F_{{\nu}{\mu}}(z)  \right) \right\} \right\rangle,
\label{risfin}
\end{eqnarray}  
where the dual tensor field is defined to be $\hat{F}^{\mu\nu} \equiv 
\varepsilon^{\mu \nu\rho\sigma} F_{\rho\sigma}/2$
Here ${\rm T_s}$ is the time-ordering prescription for spin matrices, P is the path-ordering prescription 
for gauge matrices along the loop $\Gamma$, $\Gamma_1$ denotes the path going from $(t_{\rm i}, {\bf y}_1)$
to $(t_{\rm f},{\bf x}_1)$  along the quark trajectory $(t,{\bf z}_1(t))$, $\Gamma_2$ the path going from 
$(t_{\rm f}, {\bf x}_2)$ to $(t_{\rm i}, {\bf y}_2)$  along the antiquark trajectory $(t, {\bf z}_2(t))$ 
and $\Gamma$ is the path made by $\Gamma_1$ and $\Gamma_2$  closed by the two straight lines joining 
$(t_{\rm i}, {\bf y}_2)$ with $(t_{\rm i}, {\bf y}_1)$ and $(t_{\rm f}, {\bf x}_1)$ 
with $(t_{\rm f}, {\bf x}_2)$ (see Fig. \ref{plwil2}). Finally Tr denotes the trace over the gauge matrices. 
Note that the right-hand side of (\ref{risfin}) is manifestly gauge invariant.

Defining the angular bracket term in Eq. (\ref{risfin}) as (more rigorously this should correspond 
to the leading matching between NRQCD and pNRQCD)
\begin{equation}
\left\langle \frac{1}{3}
{\rm Tr \, T_s \, P} \exp \ldots\right \rangle = {\rm T_s} \exp\left[ -i\int_{t_{\rm i}}^{t_{\rm f}} dt \,
V_{{\rm Q}\bar{{\rm Q}}}({\bf z}_1,{\bf z}_2,{\bf p}_1,{\bf p}_2,{\bf S}_1,{\bf S}_2) \right];
\label{risfin2}
\end{equation}
we obtain that 
\begin{equation}
i\frac{\partial}{\partial T} K= \left[ \sum_{j=1}^2 \left(
m_j+ \frac{{\bf p}_j^2}{2m_j}-\frac{{\bf p}_j^4}{8m_j^3}\right) +V_{{\rm Q}\bar{{\rm Q}}} \right] K,
\end{equation}
where $V_{{\rm Q}\bar{{\rm Q}}}$ is {\it the complete  QCD (quenched) quark-antiquark potential at the order $v^4$}. 
Expanding the logarithm of the left-hand side of (\ref{risfin2}) up to $1/m^2$, we find\footnote{
Notice that $\int_{\Gamma_j} dz^{\mu}f_{\mu}(z) = (-1)^{j+1} \int_{t_{\rm i}}^{t_{\rm f}} dt ( f_0(z_j) 
- {\dot{{\bf z}}}_j \cdot {\bf f} (z_j))$, where $z_j=(t,{\bf z}_j(t))$. The factor $(-1)^{j+1}$  
accounts for the fact that world line $\Gamma_2 $  runs from $t_{\rm f}$ to $t_{\rm i}$.
We also use the notation $z_j^{\prime}=(t^{\prime},{\bf z}_j(t^{\prime}))$.} 
\begin{eqnarray}
\int_{t_{\rm i}}^{t_{\rm f}} dt \, V_{{\rm Q} \bar{{\rm Q}}} &=& 
i \log \langle W(\Gamma) \rangle 
- \sum_{j=1}^2 \frac{g}{m_j} \int_{{\Gamma}_j}dz^{\mu} 
\left( S_j^l \, 
\langle\!\langle \hat{F}_{l\mu}(z) \rangle\!\rangle  \right. 
\nonumber\\
&~& \quad\quad -\frac{1}{2m_j} S_j^l \varepsilon^{lkr} p_j^k \, 
\langle\!\langle F_{\mu r}(z) \rangle\!\rangle 
- \left. \frac{1}{8m_j} \, 
\langle\!\langle D^{\nu} F_{\nu\mu}(z) \rangle\!\rangle  \right)
\nonumber\\
&~& \quad\quad - \frac{1}{2} \sum_{j,j^{\prime}=1}^2 \frac{ig^2}{m_jm_{j^{\prime}}}
{\rm T_s} \int_{{\Gamma}_j} dx^{\mu} \, \int_{{\Gamma}_{j^{\prime}}} 
dx^{\prime\sigma} \, S_j^l \, S_{j^{\prime}}^k
\nonumber\\
&~& \quad\quad \times \left( \, 
\langle\!\langle \hat{F}_{l \mu}(z) \hat{F}_{k \sigma}(z^{\prime})
\rangle\!\rangle  - \, 
\langle\!\langle \hat{F}_{l \mu}(z) \rangle\!\rangle
\, \langle\!\langle \hat{F}_{k \sigma}(z^{\prime}) \rangle\!\rangle  \right),
\label{potential}
\end{eqnarray}
where we recall that  
$$
\langle f(A) \rangle \equiv {1\over 3}{\rm Tr \>}{\rm P \>}
{\displaystyle\int {\cal D} A \, e^{iS_{\rm YM} (A)} f(A) \over \displaystyle\int {\cal D} A\,  e^{iS_{\rm YM} (A)}} 
$$
and we have introduced the vacuum expectation value in presence of quarks (i.e. of the Wilson loop)
$$
\langle\!\langle f(A) \rangle\!\rangle \equiv{\displaystyle \int {\cal D} A \, e^{iS_{\rm YM} (A)} 
{\rm Tr \>}{\rm P \>}  f(A) \exp \left[i g \displaystyle\oint_{\Gamma} dz^\mu A_\mu (z) \right] 
\over \displaystyle \int {\cal D} A\,  e^{iS_{\rm YM} (A)} {\rm Tr \>}{\rm P \>}
{\exp \left[i g \displaystyle\oint_{\Gamma} dz^\mu A_\mu (z) \right] } }.
$$

In this way one obtains from QCD the static, spin-dependent and velocity 
dependent terms that control the quarkonium spectrum and that were introduced on a pure 
phenomenological basis in Sec. 3, cf Eqs. (\ref{vfenum})--(\ref{vvel}):
\begin{equation} 
V_{{\rm Q} \bar {\rm Q}} = V_0 + V_{\rm VD} + V_{\rm SD} \> .
\end{equation}
These terms have a physical direct interpretation.

The spin independent part of the potential, $V_0 + V_{\rm VD}$, is obtained in (\ref{potential}) 
from the expansion of $\log \langle W(\Gamma) \rangle $ for 
small velocities ${\dot{\bf z}}_1(t) = {\bf p}_1/m_1$  and ${\dot{\bf z}}_2(t) = {\bf p}_2/m_2$:
\begin{equation}
i \log \langle W(\Gamma) \rangle = \int_{t_{i}}^{t_{f}} dt \, \left (V_0 (r(t)) + V_{\rm VD}({\bf r}(t))
\right ) ,
\label{willexpp}
\end{equation}
where $V_0$ is the static part and  ${\bf r}(t) \equiv {\bf z}_1(t) - {\bf z}_2 (t)$.

The spin-dependent part, $V_{\rm SD}$,  contains for each quark terms analogous to those 
one would obtained by making a Foldy--Wouthuysen transformation of a Dirac equation in an external field 
$\langle\!\langle F_{\mu \nu} \rangle\!\rangle$, along with an additional term $V_{\rm SS}$ 
having the structure of a spin-spin interaction. Therefore we can write
\begin{equation}
V_{\rm SD} = V_{\rm LS}^{\rm MAG} + V_{\rm Thomas} + V_{\rm Darwin} + V_{\rm SS} 
\label{vsd}
\end{equation}
using a notation which indicates the physical significance of the 
individual terms (MAG denotes Magnetic). The correspondence 
between (\ref{vsd}) and (\ref{potential}) is given by 
\begin{eqnarray}
\int_{t_{\rm i}}^{t_{\rm f}} dt V_{\rm LS}^{\rm MAG} \!\!&=& 
\!\!- \sum_{j=1}^2 \frac{g}{m_j} \int_{{\Gamma}_j}\!dz^{\mu} 
S_j^l \, \langle\!\langle \hat{F}_{l\mu}(z) \rangle\!\rangle  \>, \label{vmag} \\
\int_{t_{\rm i}}^{t_{\rm f}} dt V_{\rm Thomas} &=&  \!\! \sum_{j=1}^2 \frac{g}{2 m^2_j} \int_{{\Gamma}_j}\!dz^{\mu} 
S_j^l \varepsilon^{lkr} p_j^k \, \langle\!\langle F_{\mu r}(z) \rangle\!\rangle \>, \label{vthomas} \\
\int_{t_{\rm i}}^{t_{\rm f}} dt V_{\rm Darwin} &=&  \!\! \sum_{j=1}^2 \frac{g}{8 m^2_j} \int_{{\Gamma}_j}\!dz^{\mu} 
\langle\!\langle D^{\nu} F_{\nu\mu}(z) \rangle\!\rangle  \>, \label{vdarwin} \\
\int_{t_{\rm i}}^{t_{\rm f}} dt V_{\rm SS} &=&  \!\! - \frac{1}{2} \sum_{j,j^{\prime}} \frac{ig^2}{m_jm_{j^{\prime}}}
{\rm T_s} \int_{{\Gamma}_j} \!dz^{\mu} \, \int_{{\Gamma}_{j^{\prime}}} \!dz^{\prime\sigma} \, S_j^l \, S_{j^{\prime}}^k
\left( \langle\!\langle \hat{F}_{l \mu}(z) \hat{F}_{k \sigma}(z^{\prime})\rangle\!\rangle  \right. \nonumber\\
&~& \quad\quad\quad\quad\quad\quad \left. - \langle\!\langle \hat{F}_{l \mu}(z) \rangle\!\rangle
\, \langle\!\langle \hat{F}_{k \sigma}(z^{\prime}) \rangle\!\rangle  \right) \> . \label{vss}
\end{eqnarray}

If we had worked in QED, we would have obtained the same formal result. The point is that in 
QED one can calculate perturbatively the field strength  expectation values in the presence of the Wilson loop,
here one can rely on a perturbative calculation only for very short interquark  distances, shorter than the 
typical radius of the bound system.  Therefore, we have to obtain a nonperturbative evaluation of 
$\langle \!\langle F \rangle \!\rangle $ and $\langle \!\langle FF\rangle\! \rangle$ that, together with the Wilson 
loop contain, all the relevant information on the heavy quark dynamics. Notice that the expression for the 
potential contains only manifestly gauge-invariant quantities.

We have at our disposal two ways of  obtaining   the nonperturbative 
quark interaction. First, we can  exactly translate all our results in terms of field strength 
expectation values in presence of a {\it static} Wilson loop. These are plaquette insertions in 
the static Wilson loop and have been evaluated on the lattice (Bali et al. (1997)). 
The interesting  fact is that we can also perform an analytic evaluation making only an assumption on the 
nonperturbative behaviour of the Wilson loop. This is due to the fact that all the expectation values 
of Eq. (\ref{vmag})-(\ref{vss}) can be obtained as functional derivatives 
of $\log \langle W(\Gamma) \rangle $ with respect to the path, i.e. with respect to the 
quark trajectories ${\bf z}_1 (t)$ or ${\bf z}_2 (t)$. In fact let us consider the change in 
$\langle W(\Gamma) \rangle$ induced by letting $ z_j^\mu (t) \rightarrow  z_j^\mu (t) + \delta  z_j^\mu (t)$ where 
$\delta  z_j^\mu  (t_{\rm i}) = \delta  z_j^\mu (t_{\rm f}) = 0$, then we have 
\begin{equation}
g \langle\!\langle F_{\mu\nu}(z_j) \rangle\!\rangle = (-1)^{j+1}
{\delta i \log \langle W(\Gamma) \rangle \over \delta S^{\mu\nu} (z_j)} ,
\label{e20}
\end{equation}
$$ 
\delta S^{\mu\nu} (z_j) =  dz_j^\mu \delta z_j^\nu - dz_j^\nu \delta z_j^\mu ,
$$
and varying again the path 
\begin{equation}
g^2 \left(\langle\!\langle F_{\mu\nu}(z_1) F_{\lambda\rho}(z_2) \rangle\!\rangle 
- \langle\!\langle F_{\mu\nu}(z_1) \rangle\!\rangle 
  \langle\!\langle F_{\lambda\rho}(z_2) \rangle\!\rangle \right)
= - i g {\delta\over \delta S^{\lambda\rho}(z_2)} \langle\!\langle F_{\mu\nu}(z_1) \rangle\!\rangle.
\label{e21}
\end{equation}
{\it Therefore, to obtain the whole quark-antiquark potential no other assumptions are needed than  the 
behaviour of $\langle W(\Gamma) \rangle$. In particular all contributions to the spin dependent 
part of the potential can be expressed as first and second variational derivatives 
of $\log \langle W(\Gamma) \rangle$. The obtained expressions are correct up to order $v^4$. 
Higher order corrections can in principle be included systematically in the same way.}

We can use this result as a laboratory to understand confinement.
In fact any assumption on the QCD vacuum, i.e. on the nonperturbative behaviour of the Wilson loop, 
is put in direct connection, on one hand  with the lattice evaluation and on the other hand  with the 
phenomenological data. We address this issue in Sec. 6.

We conclude this section presenting one of the most common representations of the $1/m^2$ potentials
(Eichten et al. (1981) and Barchielli et al. (1988)) which we will use in Sec. 6:  
\begin{eqnarray}
V_{\rm SD} &=& 
{1\over 8} \left( {1\over m_1^2} + {1\over m_2^2} \right) 
\Delta \left[ V_0(r) +V_{\rm a}(r) \right] 
\nonumber\\
&+& \left( {1\over 2 m_1^2} {\bf L}_1 \cdot {\bf S}_1 
     - {1\over 2 m_2^2} {\bf L}_2 \cdot {\bf S}_2 \right) 
       {1\over r}  {d \over dr} \left[ V_0(r)+ 2 V_1(r) \right]
\nonumber \\
&+&
{1\over m_1 m_2}\left( {\bf L}_1 \cdot {\bf S}_2 - {\bf L}_2 \cdot {\bf S}_1 \right) 
{1\over r} {d \over dr} V_2(r) 
\nonumber\\
&+&{1\over m_1 m_2}  \left( { {\bf S}_1\cdot{\bf r} \> {\bf S}_2\cdot{\bf r}\over r^2} 
- {{\bf S}_1 \cdot {\bf S}_2 \over 3} \right) V_3(r) + {1\over 3 m_1 m_2} {\bf S}_1 \cdot {\bf S}_2 \, V_4(r) 
\label{sd}
\end{eqnarray}
with ${\bf L}_j = {\bf r} \times {\bf p}_j$
and 
\begin{eqnarray}
V_{\rm VD} &=&  \sum_{j=1}^2 {V_{\prime} (r)\over m_j} 
\nonumber\\
&+& {1\over m_1 m_2} \bigg\{ V_{\prime\prime}(r) + {\bf p}_1\cdot{\bf p}_2 V_{\rm b}(r) 
+ \left( {{\bf p}_1\cdot{\bf p}_2\over 3} - 
{{\bf p}_1\cdot {\bf r} \>~ {\bf p}_2 \cdot {\bf r} \over r^2}\right) V_{\rm c}(r) \bigg\}
\nonumber\\
&+& \sum_{j=1}^2 {1\over m_j^2}\left\{ V_{\prime\prime\prime}(r) + 
p^2_j V_{\rm d}(r) + \left( { p^2_j\over 3}  - {{\bf p}_j\cdot {\bf r} \>~ 
{\bf p}_j \cdot {\bf r} \over r^2}\right) V_{\rm e}(r) \right\} \, . 
\label{vd}
\end{eqnarray}
The brackets $\{ \cdots \}$ mean here an ordering prescription between position and momentum operators.
The functions $V_i(r)$ contain all the dynamics and are given by  expectation values 
of electric and magnetic field insertion in the static Wilson loop. 
Explicit expressions can be found in Eichten et al. (1981), Gromes (1984), Barchielli et al. (1988), (1990), 
Brambilla et al. (1990), (1993), (1997a). 

\vskip 1truecm
\leftline{\bf Exercises}
\begin{itemize}
\item[5.2.1]{ Consider a free particle in one dimension. Using Eq. (\ref{pathkappa}) 
calculate the free particle propagator. [Hint: use the discretized version of the path integral. 
See e.g. Sakurai (1985) for details].}  
\item[5.2.2]{ Demonstrate Eqs. (\ref{e20}) and (\ref{e21}) first in QED and then in QCD.}
\end{itemize}

\section{Modelling the QCD vacuum}
The limitations of a perturbative approach are appreciated when considering the vacuum. In perturbation theory the 
vacuum is approximated  as an empty state with rare quark or gluon loop fluctuations. This in turn 
means that quarks and gluons are allowed to propagate freely. We have seen that experimentally 
this is not the case. The true, nonperturbative vacuum could be better imagined as a disordered 
medium  with whirlpools of colour on different scales, thus densely populated by fluctuating fields 
whose amplitude  is so large that they cannot be described by perturbation theory.
Such a vacuum would be responsible for the fact that quark and gluons are confined. Such a vacuum would 
be responsible for the area law behaviour of the Wilson loop.

It is established that the QCD vacuum is (phenomenologically) characterized by various 
nonperturbative condensates, for a recent review see Shifman (1998). 
(An introduction to the topic has  been given at this school by Anatoly Radyushkin.) 
Half-dozen of them are  known: the gluon condensate 
$F_2 \equiv\langle {\alpha_s\over \pi}F_{\mu\nu}^a(0) F^a_{\mu\nu}(0) \rangle $,
the quark condensate $\langle \bar{q} q\rangle$, the mixed condensate  $\langle
\bar{q} \sigma_{\mu\nu} F_{\mu\nu}q\rangle $ and so on. 
Physically, the gluon condensate  measures the vacuum energy density $\varepsilon_{vac}$. Indeed, due to the 
scale anomaly of QCD, the trace of the energy-momentum tensor is given by
\begin{equation}
\theta_\mu^\mu = {\beta(\alpha_s) \over 2 \alpha_s}  F_{\mu\nu}^a(0) F^a_{\mu\nu}(0).
\label{tens}
\end{equation}
Then, in the lowest order expansion of the beta function, we get  
\begin{equation}
\varepsilon_{vac}  \equiv {1\over 4} \langle  \theta_\mu^\mu  \rangle 
\simeq -{\beta_0\over 32} \langle {\alpha_s\over \pi} F_{\mu\nu}^a(0) F^a_{\mu\nu}(0) \rangle .
\label{vac}
\end{equation}
Therefore, the nonperturbative gluon condensate shifts the vacuum energy 
downwards, making it advantageous. We remark that the negative sign in (\ref{vac}) is due to the asymptotic freedom!
The vacuum fields fluctuate and these fluctuations contribute to the vacuum energy density. 
High energy  modes of the fluctuating fields  are in the weak coupling regime  and can be dealt with 
perturbation theory as usual. The low frequency modes are responsible for the peculiar properties of the 
vacuum medium.

In the sum rule approach of Shifman, Vainshtein and Zakharov (1979) the effects caused by 
the vacuum fields are parameterized into few local vacuum condensates. 
This approach describes quite successfully the low-lying hadrons where the QCD string is not so relevant.
The underlying assumption is that the characteristic frequencies of the valence quarks 
in the bound state are larger than the characteristic scale parameter of the vacuum medium.  
In other words, the valence quark pair injected in the vacuum is assumed to perturb it only slightly. 
However, sum rules cannot tell us anything about confinement. This is due to the fact that there is
no local order parameter for confinement. But, as discussed in particular in Sec. 4.4,  
confinement manifests itself in the area law of the Wilson loop which is a nonlocal quantity\footnote{
We remark that instead  the local quark condensate is the order parameter of chiral symmetry breaking.}. 
In general {\it supplying an analytic form for the nonperturbative Wilson loop average amounts to defining 
a model of the QCD vacuum.} In the previous section we have seen that the static and the semirelativistic 
quark-antiquark interaction can be expressed in terms of the Wilson loop only.
Therefore, it is tempting to explore the confinement mechanism using the simplest case of the heavy quark 
interaction and  building analytic models of the QCD  vacuum to be tested directly on the lattice 
and on the phenomenology.

In Sec. 6.1 we present a pedagogical model, the so-called Minimal Area Law model,  
which takes seriously the lattice results and assumes that the logarithm of the Wilson loop 
is simply given by a constant (the string tension) times the minimal area enclosed by the Wilson loop. 
This model will turn out to be very instructive for three reasons: 1) it gives the feeling of how a real 
calculation is done in practice, 2) it gives at the end a concrete and non-trivial form for the QCD potential, 
3) it shows how the flux tube degrees of freedom emerge. In Sec. 6.2 we briefly summarize  
the main points underlying the issue of the dual Meissner effect as the mechanism of confinement. 
In order to give a concrete idea of how this mechanism could lead to 
quark confinement, we present the ordinary Abelian Higgs model of superconductivity. 
We also briefly explain how a mechanism similar to the Abelian Higgs model arises in QCD using 
the 't Hooft Abelian projection idea. We present some lattice results  obtained in Abelian projection 
on the interquark flux tube distribution and on the static potential. In Sec. 6.3 we discuss 
the flux tube structure at the light of some general low energy theorems.
Finally in Sec. 6.4 we only summarize the main ingredients of some other analytic models 
of the QCD vacuum, in particular Dual QCD and  the Stochastic Vacuum Model.

\subsection{Minimal Area Law Model (MAL)}
In this model (Brambilla et al. (1993)) $\langle W(\Gamma) \rangle$ is  approximated by the sum of a short range  
part given at the leading order by the perturbative gluon propagator $D_{\mu\nu}$ and a long-range part given 
by the value of the minimal area enclosed by the deformed Wilson loop of fixed contour $\Gamma$ 
(see Fig. \ref{plwil2}) plus a perimeter contribution $\cal P$:
\begin{eqnarray}
i \log \langle W (\Gamma) \rangle &=& i\log \langle W (\Gamma) \rangle^{\rm SR} 
+ i \log \langle W (\Gamma) \rangle^{\rm LR}
\nonumber\\
&=&  - \frac{4}{3} g^2 \oint_{\Gamma}dx^{\mu}_1 \oint_{\Gamma} dx^{\nu}_2 ~iD_{\mu \nu} (x_1-x_2)  
+ \sigma S_{\rm min} + {C\over 2} {\cal P}.
\label{mal}
\end{eqnarray}

Denoting by $u^{\mu}=u^{\mu}(s,t)$ the equation of a typical surface of contour $\Gamma$ ($s \in [0,1],\,
t \in [t_{\rm i},t_{\rm f}], \, u^0(s,t)=t, \, {\bf u}(1,t)= {\bf z}_1(t), 
\, {\bf u}(0,t)= {\bf z}_2(t) \,$) and defining ${\bf u}_{\rm T} \equiv {\bf u} - ({\bf u}\cdot {\bf n})~{\bf n}$ 
with ${\bf n} = (\partial {\bf u}/\partial s) |\partial {\bf u}/\partial s|^{-1}$, we can write:
\begin{eqnarray}
S_{\rm min} & = & \min \int_{t_{\rm i}}^{t_{\rm f}}
dt \, \int_0^1  ds\, \left[-\left( \frac{\partial u^{\mu}}{\partial t} \frac{\partial u_{\mu}}
{\partial t} \right) \left( \frac{\partial u^{\mu}}{\partial s}
\frac{\partial u_{\mu}}{\partial s} \right) + \left(
\frac{\partial u^{\mu}}{\partial t} \frac{\partial u_{\mu}}
{\partial s} \right)^2 \right]^{\frac{1}{2}}
\nonumber\\
& = & \min \int_{t_{\rm i}}^{t_{\rm f}} dt \, \int_0^1 ds \, \left|\frac{\partial 
{\bf u}}{\partial s } \right| \left\{ 1-\left[ \left(\frac{\partial {\bf u}}{\partial t} \right)_{\rm T} 
\right]^2 \right\}^{\frac{1}{2}},
\end{eqnarray}
which coincides with the Nambu--Goto action. Up to the relative order $1/m^2$ ($v^2$ in the velocity) 
the minimal surface  can be identified exactly  with the surface spanned by the straight-line 
joining $(t,{\bf z}_1(t))$ to $(t,{\bf z}_2(t))$  with $t_{\rm i} \le t \le t_{\rm f}$. 
The generic point of this surface is 
\begin{equation}
u^0_{\min}=t \quad \quad \quad {\bf u}_{\min} = s~{\bf z}_1(t) + (1-s)~ {\bf z}_2(t) \>,
\label{straight}
\end{equation}
with $0\leq s \leq 1$ and ${\bf z}_1(t)$ and ${\bf z}_2(t)$ being the positions of the quark and the antiquark 
at the time $t$. Then, the  expression for the minimal area at order $1 /m^2$ in the MAL turns out to be
\begin{eqnarray}
S_{\min} &=& \int_{t_{\rm i}}^{t_{\rm f}} dt \,  r \int_0^1 ds \, 
[1-(s~\dot{{\bf z}}_{1 \rm T} + (1-s)~\dot{{\bf z}}_{2 \rm T} )^2]^{\frac{1}{2}} 
\nonumber\\
&=&  \int_{t_{\rm i}}^{t_{\rm f}} dt \,  r \,
 \left[ 1-\frac{1}{6} \left(\dot{{\bf z}}_{1 \rm T}^2 
+ \dot{{\bf z}}_{2 \rm T}^2+ \dot{{\bf z}}_{1 \rm T} \cdot \dot{{\bf z}}_{2 \rm T} \right) + \cdots \, \, \right], 
\end{eqnarray}
where $r=|{\bf z}_1- {\bf z}_2|$. The perimeter term is given by
\begin{equation}
{\cal P} = \vert {\bf x}_1 - {\bf x}_2 \vert + \vert {\bf y}_1 - {\bf y}_2
\vert  + \sum_{j=1}^2 \int_{t_{\rm i}}^{t_{\rm f}} dt \sqrt {\dot{z}_j^\mu \dot{z}_{j\mu}} \>.
\label{per}
\end {equation}
In the limit of large time interval $t_{\rm f} - t_{\rm i}$ the first two terms in the right-hand 
side of Eq. (\ref{per}) can be neglected. By expanding also Eq. (\ref{per}) up to  $1/m^2$,  we obtain
\begin{eqnarray}
i \log \langle W (\Gamma) \rangle^{\rm LR} 
&=&  \int_{t_{\rm i}}^{t_{\rm f}} dt \, \sigma r \, \left[ 1-\frac{1}{6} 
\left(\dot{{\bf z}}_{1 \rm T}^2 
+ \dot{{\bf z}}_{2 \rm T}^2
+ \dot{{\bf z}}_{1 \rm T} \cdot \dot{{\bf z}}_{2 \rm T} \right) \right]
\nonumber\\
&+&  {C\over 2} \sum_{j=1}^2 \int_{t_{\rm i}}^{t_{\rm f}} dt 
\left( 1-{1\over 2} \dot{\bf z}_j\cdot \dot{\bf z}_j \right).
\end{eqnarray}

For what concerns the perturbative part in the limit of  large $t_{\rm f} - t_{\rm i}$ 
the only non-vanishing contribution to the Wilson loop is given by
\begin{equation}
i\log \langle W (\Gamma) \rangle^{\rm SR} 
= - \frac{4}{3} g^2 \int_{t_{\rm i}}^{t_{\rm f}} dt_1 
\int_{t_{\rm i}}^{t_{\rm f}} dt_2 ~{\dot z}_1^\mu(t_1) ~ {\dot z}_2^\nu(t_2) 
~iD_{\mu \nu} (z_1-z_2).
\label{sr}
\end{equation}
In the infinite time limit this expression is still gauge invariant. 
Expanding $z_2(t_2)$ around $t_1$ it is possible to evaluate explicitly 
from Eq. (\ref{sr}) the short-range potential up to a given order in the inverse of the mass 
(this was the task of  Exercise 4.2.3). Self-energy terms are neglected.

Eventually, in the MAL model the following static and velocity dependent potentials are obtained:
\begin{eqnarray}
V_0 &=& -{4\over 3} {\alpha_{\rm s} \over r} + \sigma r + C \>,
\label{v0mal}\\
V_{\rm b}(r)&=&{8\over 9} {\alpha_{\rm s} \over r} - {1\over 9}\sigma r \>, \quad \quad \,~
V_{\rm c}(r) = -{2\over 3} {\alpha_{\rm s} \over r} - {1\over 6}\sigma r \>,
\nonumber \\
V_{\rm d}(r)&=& -{1\over 9} \sigma r -{1\over 4} C \>, \quad \quad 
V_{\rm e}(r) = -{1 \over 6}\sigma r \>,
\label{vdmal}
\end{eqnarray}
which fulfill the relations (Barchielli et al. (1990))
\begin{eqnarray}
& & V_{\rm d}(r) +{1\over 2} V_{\rm b}(r) +{1\over 4} V_0(r) - 
{r\over 12} {d V_0(r)\over dr}=0 ,
\label{relvel1}\\
& &V_{\rm e}(r) +{1\over 2} V_{\rm c}(r) 
+ {r\over 4} {dV_0(r)\over dr}=0 .
\label{relvel2}
\end{eqnarray}

By evaluating the functional derivatives of the Wilson loop, as given by Eqs. (\ref{e20})-(\ref{e21}), 
we obtain also the spin-dependent potentials
\begin{eqnarray}
\Delta V_{\rm a}(r) &=& 0, \quad 
{d\over dr} V_1(r) = -\sigma, \quad 
{d\over dr} V_2(r) = {4\over 3} {\alpha_{\rm s} \over r^2}, 
\nonumber\\
V_3(r) &=&  4 {\alpha_{\rm s}\over r^3}, \quad 
V_4(r) = {32\over 3}\pi \alpha_{\rm s} \delta^3({\bf r}).
\label{vsmal}
\end{eqnarray}
These potentials reproduce the Eichten--Feinberg--Gromes results (Eichten and Feinberg (1981), Gromes (1984)) 
and fulfill the Gromes relation  
\begin{equation}
{d\over dr} \left[ V_0(r) +V_1(r)-V_2(r) \right] = 0 .
\label{grom}
\end{equation}
Notice that, as a consequence of the vanishing  of 
the long-range behaviour of the spin-spin potential $V_{\rm SS}$ and of the spin-orbit 
magnetic potential $V_{\rm LS}^{\rm MAG}$ in this model, there is no long-range contribution 
to $V_2$, $V_3$ and $V_4$.  Furthermore, $V_1$ has only a nonperturbative 
long-range contribution, which comes from the Thomas precession potential (\ref{vthomas}). 

The MAL model strictly corresponds to the Buchm\"uller picture (see Fig. \ref{pltube}) (Buchm\"uller (1982))  
where the magnetic field in the comoving  system is taken to be equal to zero. 
Let us first notice that the perimeter contributions at the $1 /m^2$ order can be simply absorbed in a 
redefinition of the quark masses $m_j \to m_j + C/2$. Then, let us consider the moving quark 
and antiquark connected by a chromoelectric flux tube and let us describe the flux tube as a string 
with  transverse velocity ${\bf v}_{\rm T}$. At the classical relativistic level the system 
is described by the  flux tube Lagrangian (Olsson et al. (1993))
\begin{equation}
L = - \sum_{j=1}^2 m_j \sqrt{1- {\bf v}_j^2} 
- \sigma \int_0^r dr^\prime \sqrt{1- {\bf v}_{\rm T}^{\prime 2}} \>,
\end{equation}
with $ {\bf v}_{\rm T}^\prime  = {\bf v}_{1{\rm T}}~ {r^\prime / r}  
+ {\bf v}_{2{\rm T}} (1- {r^\prime / r} ), \quad 0< r^\prime < r $.
The semirelativistic limit of this Lagrangian gives back the nonperturbative part 
of the $V_0$ and $V_{\rm VD}$ potentials in the MAL model (\ref{v0mal}) (\ref{vdmal}) 
(notice that the minimal area law in the straight-line approximation is the configuration 
given by a straight flux tube). The remarkable characteristics of the obtained $V_{\rm VD}$ potential 
is the fact that it is proportional to the square of the angular momentum and so takes 
into account the energy and the angular momentum of the string 
\begin{equation}
V_{\rm VD}^{\rm LR} = - {1\over 12 m_1 m_2}  {\sigma \over r}
({\bf L}_1 \cdot {\bf L}_2 + {\bf L}_2 \cdot {\bf L}_1 )
- \sum_{j=1}^2 { 1\over 6 m_j^2 } {\sigma \over r} {\bf L}_j^2 \>.
\label{vdflux}
\end{equation}
Finally, the nonperturbative spin-dependent part of the potential in this intuitive 
flux tube picture simply comes from the Buchm\"uller ansatz that the  chromomagnetic 
field is zero in the comoving framework of the flux tube.

We notice that even if $V_1$ seems to arise from an effective Bethe--Salpeter kernel 
which is a scalar and depends only on the momentum transfer, a simple convolution 
kernel (see Eq. (\ref{kernel})) cannot reproduce the correct velocity dependent potential 
(\ref{vdflux}) or equivalently (\ref{vdmal}). Nevertheless the behaviour (\ref{vdflux}) 
is important to reproduce the spectrum\footnote{The extension of the Wilson loop approach 
to the relativistic treatment of heavy-light bound states supports the fact  that the nonperturbative 
kernel {\it is not} a scalar convolution kernel, cf. Brambilla (1998) and Brambilla et al. (1997b).}.

The static, spin-dependent and velocity-dependent potentials have been recently evaluated 
on the lattice (Bali et al. (1997)) and, at the present level of accuracy, confirm the prediction 
of this simple model.

\subsection{Dual Meissner effect: a simple example}
\begin{figure}[htb]
\vskip -0.2truecm
\makebox[1.0truecm]{\phantom b}
\centerline{\epsfxsize=8truecm\epsffile{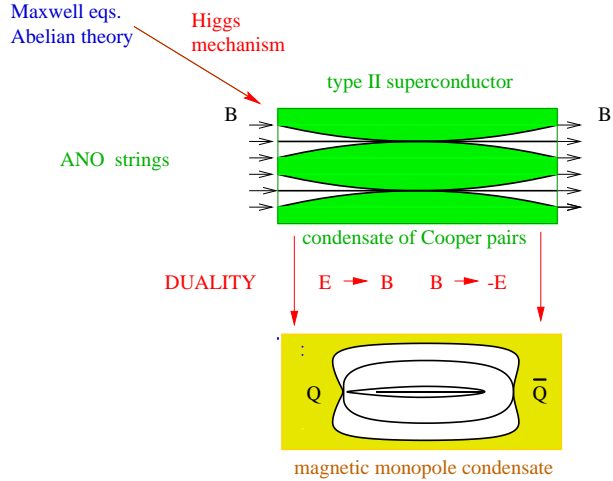}}
\vskip 0.2truecm
\caption{\it  The QCD vacuum as a ``dual'' superconductor .} 
\label{pldual}
\end{figure}

In the previous sections we have seen that the main characteristic of the quark nonperturbative 
interaction is the chromoelectric flux tube formation (see Fig. \ref{plac}).
The formation of a flux tube is reminiscent of the Meissner effect in (ordinary) superconductivity
(for a review see Weinberg (1986)).
As it is well-known, the superconducting media do not tolerate the magnetic field. If one imposes a certain 
flux of magnetic field through such a medium, the magnetic field will be squeezed into a thin tube 
carrying all the magnetic flux. The superconducting phase is destroyed inside the tube.
A well-known example of the string-like solution of the classical equations of motion is given 
by the Abrikosov string (Abrikosov (1957)). Superconductivity is caused by condensation of the 
Cooper pairs (pairs of electric charges). If there were monopoles and antimonopoles in
 the superconducting medium, 
then a  string would be formed and confine them. Therefore, to explain the confinement 
of electric charges (the quarks), we need a condensate of (chromo)magnetic monopoles (see Fig. \ref{pldual}). 
This is the simple qualitative idea suggested by Nambu (1974) and 't Hooft (1976) and Mandelstam (1976).
Then, the vacuum of QCD behaves like a dual superconductor, where the word ``dual'' here means 
that the role of electric and magnetic quantities is interchanged with respect to an ordinary 
superconductor  (see Fig. \ref{pldual}). In the next section we present a concrete model for a superconductor.

\subsubsection{Abelian Higgs Model}
A relativistic version of the Ginzburg--Landau model for superconductivity is the Abelian Higgs model:
\begin{equation}
L = -\frac{1}{4} F^{\mu\nu} F_{\mu\nu} + {1\over 2}(D^\mu\Phi)^\dagger(D_\mu\Phi) - U(\Phi)
\label{eq:2.1}
\end{equation}
where $\Phi$ is the Higgs field, in this case a complex (charged) scalar field describing Cooper 
pairs (i.e. condensate of electrons pairs in a lattice of positive ions, a superconducting solid) 
of charge $q= 2 e$, $e$ is the electron charge. $D_\mu\Phi = (\partial_\mu - i q A_\mu)\Phi$ 
is the covariant derivative, and $U(\Phi) = \displaystyle\frac{\lambda^\prime}{4}
(\Phi^\dagger\Phi -\mu^2)^2$ is the Higgs potential. In Nielsen and Olesen (1973) it was shown that 
the Lagrangian (\ref{eq:2.1}) allows  for vortex-line solutions. These vortex-lines solutions
were approximately identified with the Nambu string.

The Lagrangian $L$ is invariant under local $U(1)$ transformations. However, if $\mu^2 >0$, 
there is a  condensation of Cooper pairs signalled by the vacuum expectation value 
$\langle\Phi\rangle\neq 0$. This corresponds to the spontaneous breaking of the U(1) electric 
symmetry\footnote{This is often referred as spontaneous breaking of the local gauge symmetry. 
However, this statement is somewhat inaccurate. In this kind of theories the vacuum {\it does not}  
break local gauge invariance. Any state in the Hilbert space that fails to be invariant  under local gauge 
transformation is  an unphysical state. The vacuum is entirely gauge invariant at variance with what happens in 
theories  with a global symmetry. See 't Hooft (1994).}. As a consequence the photon 
becomes massive inside a superconducting body and an external magnetic field can penetrate 
the body only to a finite depth $\lambda$ equal to the inverse of the photon mass (Meissner effect). 
In fact putting $\Phi = \rho e^{i\theta}$, $\rho>0$, and  $\tilde A_\mu = (A_\mu - \partial_\mu\theta)$, 
$L$ can be rewritten  in the gauge-invariant form
\begin{equation}
L = -\frac{1}{4} F^{\mu\nu} F_{\mu\nu} + \frac{M^2}{2}\tilde A^\mu\tilde A_\mu + {\tilde{L}} [\rho]
\label{eq:2.4}
\end{equation}
where ${\tilde{L}}$ is the sector of the Lagrangian describing the propagation of the massive 
field $\rho$ (mass $M_\phi$). The equations of motion for the electromagnetic field read
\begin{equation}
\partial^\mu F_{\mu\nu} + M^2 \tilde A_\nu = 0
\label{eq:2.5}
\end{equation}
with $M^2 = q^2 \langle\Phi\rangle^2$, the mass acquired by the photon.      

In a stationary state with no charges $A_0=0$, $\partial_0 {\bf A }= 0$, Eq. (\ref{eq:2.5}) gives
$({\bf H} = \nabla\wedge{\bf A})$
\begin{eqnarray}
\nabla\wedge{\bf H} + M^2 {\tilde {\bf A}} &=& 0 
\label{eq:2.6a}\\
\nabla^2 {\bf H} + M^2{\bf H} &=& 0
\label{eq:2.6b}
\end{eqnarray}
Eq. (\ref{eq:2.6a}) means that a permanent current (London current) ${\bf j} = M^2
{\tilde {\bf A}}$ exists, and since ${\bf E} = 0$ and ${\bf E} = \sigma_c \,{\bf j}$ ($\sigma_c$ 
is the conductivity), $\sigma_c=0$. 

The key parameter is the order parameter $\langle\Phi\rangle$ which signals the Higgs phenomenon.
There are two characteristic lengths in the system related to $\langle\Phi\rangle$: the correlation length 
of the $\Phi$ field (or the inverse Higgs mass) $\Lambda = 1/M_\phi$, 
$M_\phi=\lambda^\prime \langle \Phi \rangle$, and the penetration depth of the photon, 
$\lambda = 1/M$. If $\lambda > \Lambda$ the superconductor is called of type II, and the formation of Abrikosov 
flux tubes is favored in the process of penetrating the material with a magnetic field. If the opposite inequality 
holds, $\lambda < \Lambda$, it happens that, when the magnetic field is increased,  there is an abrupt
penetration of it at some value and superconductivity is destroyed. The superconductor is of type I.

Summarizing, in this model the condensation of electric charges leads to the formation of a quantized 
flux tubes (see Exercise 6.2.1.1) whose radius and shape are controlled by $\Lambda$ and $\lambda$.  
If a magnetic monopole and antimonopole were introduced into such a superconducting medium 
they would be connected by a flux tube of finite energy per unit length (finite string tension). 
Thus magnetic monopoles would be  confined due to a linear potential.  

\vskip 1truecm
\leftline{\bf Exercises}
\begin{itemize}
\item[6.2.1.1]{ Show that a side consequence of the Meissner effect is the flux quantization. 
[Hint: Calculate the integral $\displaystyle \oint \tilde{\bf A} \cdot d {\bf x}$  
around a large circle centered on the flux tube.] } 
\item[6.2.1.2]{ Consider the Maxwell equations in a relativistic medium without sour\-ces. Instead of 
expressing as usual the ${\bf E}$ and ${\bf B}$ fields in terms of the magnetic potential 
$A_\mu$, express the ${\bf D}= \varepsilon {\bf E} $ and ${\bf H}={\bf B}/\mu$ fields in terms 
of a (dual) electric potential $C_\mu$. Write down the Maxwell equations in a covariant form in terms 
of the field strength tensor of $C_\mu$. How do you have to modify the definition of this 
field strength tensor in the presence of sources?}
\end{itemize}

\begin{figure}[htb]
\vskip 0.1truecm
\makebox[2.5truecm]{\phantom b}
\epsfxsize=8truecm\epsffile{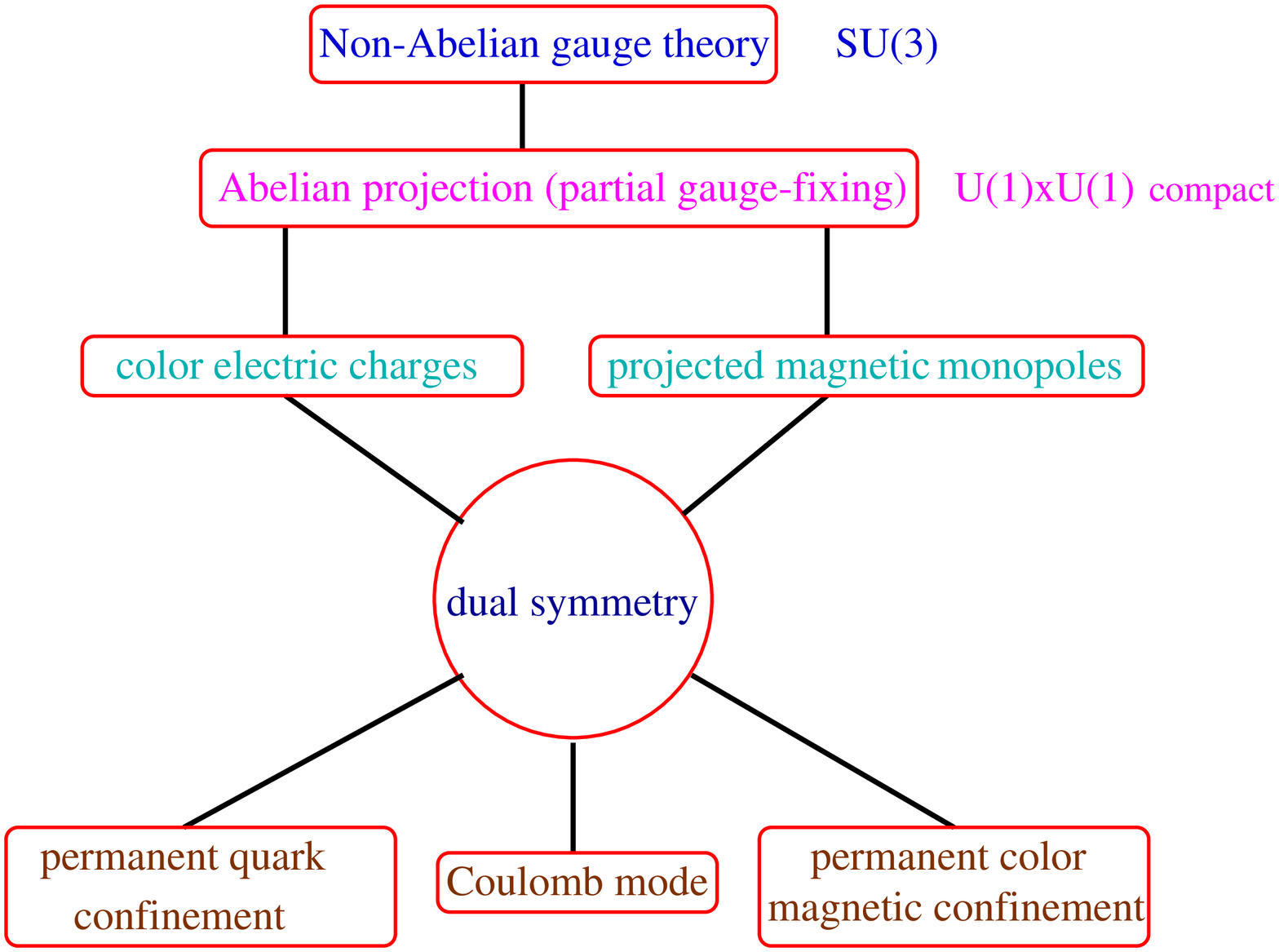}
\vskip 0.2truecm
\caption{\it  Schematic diagram illustrating  how a Non-Abelian gauge theory 
reduces to an Abelian theory with electric charges and monopoles and  the pattern of vacuum 
phases (see Sec. 4.4).} 
\label{plhooft}
\end{figure}

\subsubsection{The QCD Vacuum}
We have seen in the previous section how in ordinary superconductivity the condensation of electric
charges gives origin to  monopole confinement. On the other hand, in all the theories allowing for an
analytic proof of electric charge confinement, like compact electrodynamics, the Georgi--Glashow model or 
some supersymmetric Yang--Mills theories (see Alvarez-Gaum\'e (1997)), this is due to the condensation of monopoles.
However, if we want to apply this idea to QCD we have first to understand in which way we can obtain 
Abelian degrees of freedom and monopoles in QCD, since {\it in the QCD Lagrangian there are no Higgs fields!}.
Subsequently we have to prove that the monopoles actually do  condense. The application of this idea to  non-Abelian 
gauge theories is based on the so-called Abelian projection, see 't Hooft (1981). The Abelian projection 
is a partial gauge fixing (of the off-diagonal components of the gauge field) under which the Abelian 
degrees of freedom remain unfixed. In QCD the Abelian projection reduces the $SU(3)$  gauge symmetry to a $U(1)^2$ 
gauge symmetry. The Abelian projection monopoles  appear as topological quantities in the residual Abelian 
channel. QCD  is then reduced to an Abelian theory with electric charges and monopoles, see Fig. \ref{plhooft}.
Precisely, it can be regarded  as an Abelian gauge theory  with magnetic monopoles and charged 
matter fields (quarks and off-diagonal gluons). Then, the dual superconductor picture is 
realized if these Abelian monopoles condense: this  causes confinement of the particles that are electrically 
charged with respect to the above ``photons''. In this scenario large distance (low momentum) properties 
of QCD are carried by the Abelian degrees of freedom (Abelian dominance) and specifically 
by the monopole configurations (monopole dominance), see Suzuki (1993).

We do not have enough space to give further details. The reader is referred to Ex. 6.2.2.1 and to the 
reviews e.g. of Bali (1998), Chernodub et al. (1997), Di Giacomo (1998), Haymaker (1998). 
In the last years intensive  work has been done to collect information on the quark confinement mechanism  
via lattice measurements. Depending on the picture, the excitations giving rise to confinement  are thought to be 
magnetic monopoles, instantons, dyons, centre vortices (see Faber (1998)), etc. The above ideas 
are not completely disjoint and do not necessarily exclude each other. The above mentioned topological 
excitations indeed, are found to be correlated  with each other in the lattice studies (e.g. correlations 
between instantons and monopoles). Many questions are still not completely settled. 
Here, we would like to summarize  briefly the present understanding and to show some 
of the lattice measurements made in Abelian projection.

\begin{figure}[htb]
\vskip 0.2truecm
\makebox[1.0truecm]{\phantom b}
\centerline{
\epsfxsize=8truecm
\epsffile{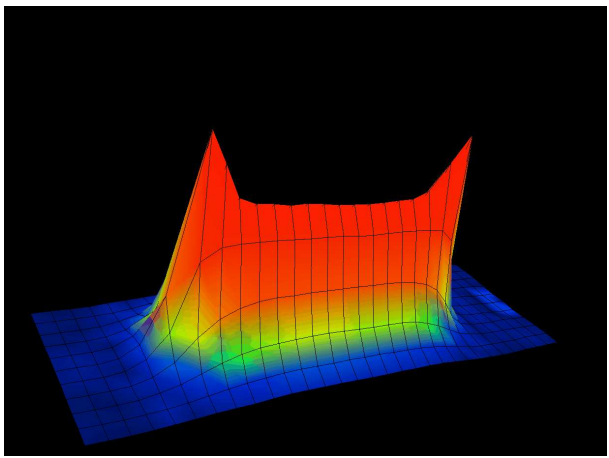}}
\vskip 0.2truecm
\caption{\it Action density distribution in Maximal Abelian Projection of $SU(2)$ measured on a lattice $V=32^4$ 
at $\beta=2.5115$. The physical quark-antiquark distance is $1.2$ fm. Figure provided by G. Bali.}
\label{plpac}
\end{figure}

First, we mention some general problems one encounters in this approach. 
1) The identification of ``photon fields'' and monopoles is a gauge  invariant process.
However, the choice of the operator that defines the Abelian projection is somehow ambiguous.
There are many different ways of making the Abelian projection. The physics, e.g. the monopole condensation, 
should be independent of the gauge fixing. 2) The actual composition of the condensate should be known. 
3) The origin of the monopole potential that yields to a non-vanishing 
vacuum expectation value of the magnetic condensate should be understood.

On the other hand it is clear that: 1) monopoles are condensed in the confinement phase and this 
independently of the choice of the Abelian projection. 2) Monopoles are 
responsible for the main part of the nonperturbative dynamics.
In particular Abelian and monopoles dominance seem to hold (in the Maximal Abelian Projection
(MAP)): expectation values of physical quantities in the non-Abelian theory coincide with  
(or are very close to) the expectation values of the corresponding Abelian operators 
in the Abelian theory obtained via Abelian projection. In other words, disregarding  the off-diagonal gluons, 
the long range features of a $SU(N)$ gauge theory are reproduced while short range features may be altered.
Monopole dominance means that the same result  can be approximately calculated  in terms only 
of the monopole currents extracted from the Abelian fields. 

\begin{figure}[htb]
\vskip 0.2truecm
\makebox[1.0truecm]{\phantom b}
\centerline{
\epsfxsize=8truecm
\epsffile{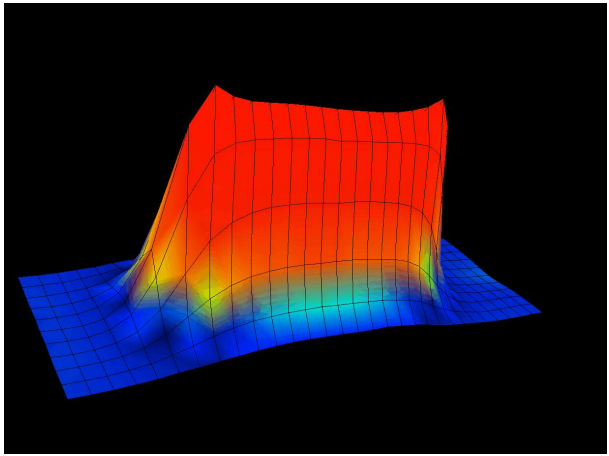}}
\vskip 0.2truecm
\caption{\it Monopole contribution to the action density distribution in 
Maximal Abelian Projection of $SU(2)$. Parameters as in Fig. \ref{plpac}. }
\label{plpmon}
\end{figure}

As a general conclusion one can say that {\it the lattice data in MAP show that the lattice gluodynamics 
is in a sense equivalent at large distances to the Dual Abelian Higgs model at the border between 
superconductor of type I and superconductor of type II}. More precisely  
the Abelian monopole action has been extracted from $SU(2)$ lattice data in MAP
and mapped into an Abelian Higgs model. From this the effective string theory has even been reconstructed 
and it occurs that the classical string tension of the string model is close to the quantum string tension
of $SU(2)$ lattice gluodynamics, cf. Chernodub et al. (1999). The field distributions in MAP are found to satisfy 
the dual Ginzburg Landau equations and the coherence and the penetration lengths have been measured, cf.  Bali (1998).

\begin{figure}[htb]
\vskip 0.1truecm
\makebox[1.0truecm]{\phantom b}
\centerline{
\epsfxsize=8truecm
\epsffile{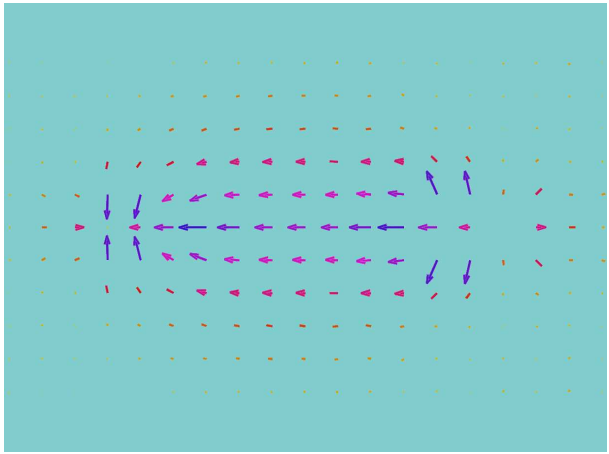}}
\vskip 0.5truecm
\caption{\it  Electric field at 12 lattice units between two static quarks in $SU(2)$ 
measured on a lattice $V=32^4$ at $\beta=2.5115$. The lattice spacing is fixed on 
$\sigma= (440$ MeV$)^2$. The physical quark-antiquark distance is 1 fm. Figure provided by G. Bali.}
\label{ple}
\end{figure}

In Fig. \ref{plpac}  we present the action density distribution between two static quarks in $SU(2)$ in MAP 
to be compared on one hand with the action density in Fig. \ref{plac} and on the other hand with the monopole 
contribution in Fig. \ref{plpmon}. We see that still the ``photon'' (neutral gluon) contributes to the self-energy 
of charges: the monopole part is free of self-energy while the ``photon'' part is free of string tension.
Electric flux tubes (cf. Fig. \ref{ple}) are found to be significantly thinner after Abelian projection. 

\begin{figure}[htb]
\vskip 0.2truecm
\makebox[1.0truecm]{\phantom b}
\centerline{\epsfxsize=8truecm\epsffile{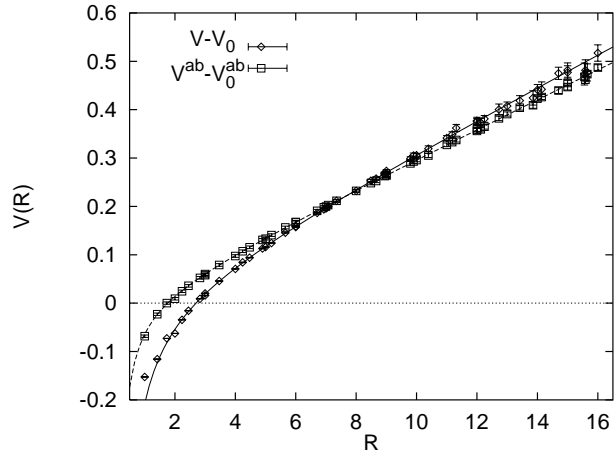}}
\vskip 0.2truecm
\caption{\it Static potential in $SU(2)$ (diamonds) and in the Abelian projection of $SU(2)$
 (squares) (in lattice units, $a\simeq 0.081$ fm). Bali et al. (1996).} 
\label{plppot1}
\end{figure}

\begin{figure}[htb]
\vskip 0.2truecm
\makebox[1.0truecm]{\phantom b}
\centerline{\epsfxsize=8truecm\epsffile{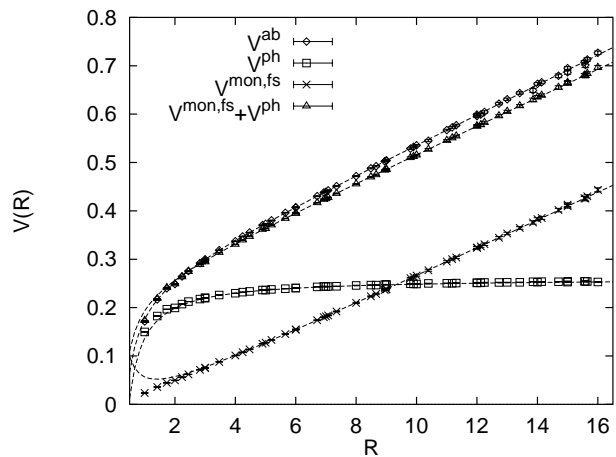}}
\vskip 0.2truecm
\caption{\it The Abelian projected $SU(2)$ potential (diamonds) in comparison with the 
``photon'' contribution (squares), the monopole contribution (crosses) and the sum of the two parts (triangles) 
(in lattice units, $a=0.081$ fm). No self-energy constant have been subtracted. Bali et al. (1996).} 
\label{plppot2}
\end{figure}

At the end of this section we present two plots of the static potential. Fig. \ref{plppot1} is a plot of 
the static potential calculated by means of either non-Abelian or Abelian projected Wilson loops. If one takes into 
account that in the Abelian projected case all the ``charged  components'' of $A_\mu$ are neglected, one finds an 
impressive agreement between the two curves. Fig. \ref{plppot2} shows the static potential 
calculated in the Abelian projection. The  photon contribution and  the monopole contribution have been 
calculated separately. This plot confirms the action density result discussed above. Moreover, it is possible 
to quantify $\sigma_{U(1)}\simeq 92 \% \, \sigma_{SU(2)}$ and  $\sigma_{mon} \simeq 95 \% \, \sigma_{U(1)}$.
Notice that in the Abelian projection the coefficient of the Coulombic part comes out to be 
smaller by more than a factor of two.

{\it The original phenomenological Cornell potential that was confirmed  by the lattice simulations  
on the static Wilson loop in Sec. 4.5, is now understood in terms of monopole contributions  while   
 the  monopole currents are the origin of the chromoelectric flux tube.}

It is also possible to construct infrared effective ``analytic'' models  based upon the assumptions 
and on the results presented above and to use them to explain the QCD low energy physics. 
As already discussed, in the Wilson loop formalism only an assumption on the Wilson loop behaviour 
is necessary. This is outlined in the next sections.

\vskip 1truecm
\leftline{\bf Exercises}
\begin{itemize}
\item[6.2.2.1]{ This is a simple example of Abelian projection for SU(2). Consider the following condition
$F_{12}(x) = \hbox{diagonal matrix}$.  Show that $F_{12}$ is still invariant under the 
U(1) gauge transformation $\Omega(x)_{ij}$ defined as  
$$
\Omega_{12}=\Omega_{21}=0  \qquad \Omega_{11}=\Omega_{22}^* = e^{i \alpha(x)} 
\qquad \alpha \in [0,2\pi).  
$$
Define $A_\mu^{\pm}\equiv A_\mu^1\pm A_\mu^2$, $1,2,3=$ colour indices. Show that  $A_\mu^3 $ transforms as a photon 
under $\Omega$ while $A_\mu^\pm$ transform as charged fields. Define an Abelian field strength tensor 
in terms of $A_\mu^3$ and show that, if the matrix of the gauge transformation $\Omega$ contains 
singularities, also the Abelian field strength tensor may contain singularities (monopoles).}
\end{itemize}

\subsection{Low energy theorems and flux tube}
We have seen that one of the relevant features of confinement is the interquark flux tube formation,  
see Figs. \ref{plac}, \ref{ple} in $SU(2)$ and Figs. \ref{plpac}, \ref{plpmon} in the MAP of $SU(2)$. 
Typical quantities of interests are the transverse extent of the flux tube  and the nature of 
the colour fields (i.e. electric or magnetic), see Bali et al. (1995) and Green et al. (1997).
There are two low energy theorems relating  the potential of a static quark-antiquark 
pair  with the total energy and the action stored in the flux tube between the sources   
(in lattice QCD these are know as Michael's sum rules (Michael (1987)), Dosch et al. (1995), 
Novikov et al. (1981), Shifman (1998), 
\begin{eqnarray}
V_0(r) &=& {1\over 2} \langle \int d^3x (-{\bf E}(x)^2 +{\bf B}(x)^2)\rangle_{r\times T}\nonumber\\
V_0(r)+ r{\partial V_0(r)\over \partial r} &=& {1\over 2} {\beta({\alpha_{\rm s}})\over \alpha_{\rm s}}
\langle \int d^3x ({\bf E}(x)^2 +{\bf B}(x)^2)\rangle_{r\times T}.  
\label{teor}
\end{eqnarray}
$\langle  \cdots \rangle_{r\times T}$ denotes the expectation value in the presence of the 
static Wilson loop where the expectation value in the absence of the sources has been subtracted. 
The formulas  are in Euclidean space and renormalized composite operators are used.
The squared electric and magnetic field strengths are separately not renormalization group invariant.  
Thus statements like $\langle g^2 {\bf B}^2\rangle_{r\times T} \simeq 0$ or 
$\langle g^2 {\bf E}_\perp^2\rangle_{r\times T} \simeq 0$ are scale dependent.
As we already pointed out, on the lattice the averages of the squared components 
of the colour fields are found to be roughly equal (i.e. $\langle E_i^2\rangle_{r\times T}$  $\simeq$  
$\langle B^2_j\rangle_{r\times T}$, $i,j=1,2,3$), while in the greater part of the models the flux tube 
is mainly made of the longitudinal electric field. Eqs. (\ref{teor}) fixes the scale at which each 
of these situations can be fulfilled. In particular for models, typically describing the long range 
behaviour with 
$V_0 \simeq \sigma r$, one has $\beta/\alpha_{\rm s} \simeq -2$. Taking the three-loop beta 
function, this gives $\alpha_{\rm s} \simeq 0.6$.

\subsection{Models of the QCD Vacuum}
We  briefly mention some analytic models of the QCD vacuum. 
We refer the reader to the papers quoted below  for further details.

\begin{figure}[htb]
\vskip 0.2truecm
\makebox[1.0truecm]{\phantom b}
\centerline{\epsfxsize=7truecm\epsffile{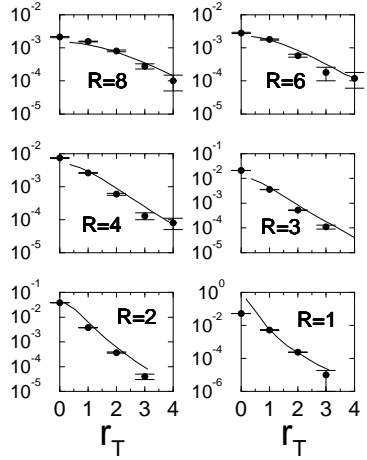}}
\vskip -0.4truecm
\caption{\it On a semi-log scale, a comparison of the total energy profile for dual QCD (solid line)
and in the lattice predictions (solid circles) for $R$ (=$q\bar{q}$ distance) ranging from 8 down to 1 lattice units 
as a function of the transverse coordinate $r_{\rm T}$ (with respect to the interquark string). 
All profiles are in lattice units with $a\simeq 0.6$ $Gev^{-1}$. From Green et al. (1997).}
\label{bbzact}
\end{figure}

\subsubsection{Dual QCD (DQCD)}
Impressive progresses have been done recently towards an understanding of confinement 
via duality and monopoles condensation in supersymmetric theories, see e.g. Alvarez-Gaum\'e et al. (1997).
However, it is also possible to construct a dual  effective theory  of long distance 
Yang--Mills theory, Baker et al. (1986-1996), see also Maedan et al. (1989). We call this theory Dual QCD (DQCD).
 
DQCD  is a concrete realization of the Mandelstam and 't Hooft dual superconductor mechanism of confinement.
It describes the QCD vacuum as a dual superconductor on the border between type I and II. 
The Wilson loop approach supplies a simple method to connect 
averaged local quantities in QCD and in dual QCD. For large loops we assume
\begin{equation}
\langle W(\Gamma) \rangle \simeq  {\displaystyle{\int \!{\cal D} { C} \,  {\cal D} {B} \,
\exp \left[ i \int dx  L(G_{\mu\nu}^{\rm S}) \right] }
\over \displaystyle{\int \!{\cal D} { C}\, {\cal D} { B} \, \exp \left[ i \int dx  L(0) \right]  }}
\label{deq}
\end{equation}
where $L$ is the effective dual Lagrangian. The fundamental variables are an octet of dual potential $C_\mu$  
coupled minimally to three octets of scalar Higgs field $B_i$ carrying magnetic colour charge,
the dual coupling constant being $e = 2\pi/g$. Notice that the dual transformation exchange
the strong coupling limit of QCD with the weak coupling limit of DQCD.
The monopole fields $B_i$ develop nonvanishing expectation values $B_{0i}$ giving rise to massive $C_\mu$ 
and to a dual Meissner effect. Dual potentials couple to electric colour charges like ordinary potentials 
couple to monopoles, i.e. $C_\mu$ couple to the quark-antiquark pair via a Dirac string connecting  the pair.
It turns out that, for the description of a quark-antiquark state, the relevant subset 
of the theory is a (dual) Abelian Higgs model. In this case indeed, the effective Lagrangian is 
explicitly given by ${ L}(G_{\mu\nu}^{\rm S}) = 2~{\rm Tr} \{ - {1\over 4} {G}^{\mu\nu} {G}_{\mu\nu}$ 
$+ {1\over 2} ({D}_\mu {B}_i)^2  \} - U({B}_i)$, where $G_{\mu\nu}=(\partial_\mu C_\nu-\partial_\nu
C_\mu + G^S_{\mu\nu})$. The dual field is directed along the hypercharge matrix $Y$
in the colour space, $G_{\mu\nu}^{\rm S}(x)  = g \,\epsilon_{\mu\nu\alpha\beta} \displaystyle\int ds \int d\tau\, 
\displaystyle{\partial y^\alpha \over \partial s} {\partial y^\beta\over \partial \tau} \delta(x-y(s,\tau))\, Y$,
($y(s,\tau)$ is a  world sheet with boundary $\Gamma$ swept out by the Dirac string)  
and $U({B}_i)$ is  the Higgs potential with the minimum values $B_{0i}$ chosen in order to completely break 
the dual $SU(3)$ symmetry. It essentially coincides with the Abelian Higgs Lagrangian (plus sources)
of Eq. (\ref{eq:2.1}) where the fields $A_\mu$ play the role of the dual potentials $C_\mu$ 
and a combination of fields $B_i$ plays the role of the Higgs field $\Phi$. 

From (\ref{deq}) it follows that (see Baker et al. (1996))
\begin{equation} 
\langle\!\langle F_{\mu \nu} (z_j)\rangle\!\rangle = {2\over 3}  \, 
\varepsilon_{\mu\nu \rho\sigma}\langle\!\langle G^{\rho\sigma} (z_j) \rangle\!\rangle_{Dual}\, .
\label{valdual}
\end{equation}
Therefore, it is possible to relate averaged values of local quantities in QCD and in the dual theory.
The nonperturbative parameters are  the  v.e.v. of the Higgs field and the  
coupling constant of the Higgs potential  (from these the penetration length and the correlation length can 
be constructed).  Flux tube configurations with finite $r_{MS} \sim {1/ M}$, $M=$ mass of the dual gluon, 
arise (Baker et al. (1996)) from the numerical solution of the classical
dual Ginzburg--Landau type of equations obtained from $L$. The flux tube profile agrees 
well with lattice data (Green et al. (1996)), see Fig. \ref{bbzact}. 
It is the presence of the Higgs field which confines the transverse  energy distribution in a flux tube.
The string tension $\sigma$ comes from the integral of the   exponentially decreasing energy distribution.

\begin{figure}[htb]
\vskip 0.2truecm
\makebox[1.0truecm]{\phantom b}
\centerline{\epsfxsize=6truecm\epsffile{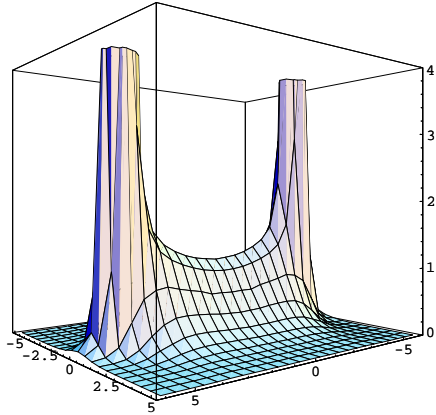}}
\vskip -0.4truecm
\caption{\it 
The energy density distribution in GeV/$fm^3$ caused by the nonperturbative correlator $D$ and the colour Coulomb 
contribution $D_1$ with $\alpha_s=0.57$. In the figure $a=T_g=0.35 fm$, the $q\bar{q}$ distance is
9 $T_g$. From R\"uter et al. (1995).}
\label{rut}
\end{figure}

\subsubsection{Stochastic Vacuum Model}
The Stochastic Vacuum Model (Dosch (1987), Dosch et al. (1988)) is based on the idea that 
low frequency contributions in the functional integral can be taken into account by 
a simple stochastic process with a converging cluster expansion. 
Under this assumption the Wilson loop manifests an area law behaviour and therefore linear confinement. 
The model  does not give rise to confinement for an Abelian gauge theory.
In order to make  quantitative predictions it is convenient to make a more radical assumption, namely 
that all higher  cumulants can be neglected  as compared to the two point function.
This assumption  appears to be in agreement with a recent lattice analysis (Bali et al. (1998)).
Then, the stochastic process is Gaussian and all fields correlators are reduced to products of two point 
functions by factorization. This means that the Wilson loop can be approximated  simply by (in Euclidean space) 
\begin{equation}
\langle W(\Gamma) \rangle \simeq  
\exp\left[ - \displaystyle {g^2\over 2} \int_{S(\Gamma)} \!\!\!\!d S_{\mu\nu}(x) 
\int_{S(\Gamma)} \!\!\!\!d S_{\lambda\rho}(y) 
\langle  F_{\mu\nu}(x) U(x,y) F_{\lambda\rho}(y) U(y,x) \rangle \right]
\label{Wsvm}
\end{equation}
where $S(\Gamma)$ is a surface with contour $\Gamma$, typically the minimal area surface. 
The nonperturbative dynamics is given in terms of one unknown function only: 
the non-local gluon condensate 
\begin{eqnarray}
&& g^2 
\langle U(0,x) F_{\mu\nu}(x) U(x,0) F_{\lambda\rho} (0) \rangle =
\bigg\{ (\delta_{\mu\lambda}\delta_{\nu\rho} - 
\delta_{\mu\rho}\delta_{\nu\lambda})(D(x^2) + D_1(x^2)) 
\nonumber \\
&&  + (x_\mu x_\lambda \delta_{\nu\rho} - 
x_\mu x_\rho \delta_{\nu\lambda} 
+ x_\nu x_\rho \delta_{\mu\lambda} - x_\nu x_\lambda \delta_{\mu\rho})
{d\over dx^2}D_1(x^2) \bigg\} .
\label{due}
\end{eqnarray}
Eq. (\ref{due}) is the most general Lorentz decomposition of the two-point correlator. 
The dynamics is contained in the form factors $D$ and $D_1$. 
The function $D$ is responsible for the area law and confinement.

Lattice simulations (for a sum-rule analysis see Dosch, Eidem\"uller and Jamin (1998))
show that the $D$ and $D_1$ functions  exhibit a long-range exponential fall off 
$\simeq F_2 \exp\{-\vert x\vert/T_g\}$ where the correlation length $T_g$ is about 0.2 $\div$ 0.3 fm
(D'Elia et al. (1997)). The model has thus two nonperturbative parameters $F_2$ and $T_g$.  
The static potential is given by 
\begin{equation}
V_0(r) \simeq  F_2 \int_0^\infty d\tau \int_0^r d \lambda\, (r-\lambda) D(\tau^2 +\lambda^2)
\label{svpot} 
\end{equation}
and the string tension  $\sigma$ emerges as an integral on the function $D$,  
$\sigma \simeq F_2 \displaystyle \int_0^\infty d\tau $
$\displaystyle \int_0^\infty  d\lambda$  $D(\tau^2 +\lambda^2)$, in the limit ${T_g/ r}\to 0$.
The field distribution between the quark and the antiquark is a flux tube (R\"uter et al. (1995)) 
with  $r_{MS} \simeq 1.8\, T_g$, see Fig. \ref{rut}. We refer to  Dosch (1994), Nachtmann (1996) 
and Simonov (1996) for a complete description of the model and its applications.

\vskip 0.7truecm
Finally, we mention in this neither exhaustive nor complete list, the flux tube model of Isgur et al.
(1983). This model is extracted from the strong coupling limit of the QCD lattice Hamiltonian. 
A $N$-body discrete string-like Hamiltonian describes the gluonic degrees of freedom. The limit 
$N\to \infty$ corresponds to a localized string with an infinite number of degrees of freedom. 

All the mentioned   models allow to obtain via Eqs. (\ref{potential}),  (\ref{e20}),  (\ref{e21}), 
the complete semirelativistic quark interaction. For a discussion and a comparison of the result 
see Baker et al. (1997), Brambilla et al. (1997a) and Brambilla (1998).  
A relation between DQCD and SVM has been established in Baker et al. (1998).

Here, we only stress that we need two parameters like $T_g$ and $F_2$ 
to control the structure of the flux tube. Had we only one parameter, like the string tension 
$\sigma$, we could only encode the information of a constant energy density in the flux tube.
However, the whole structure is important and also the information about the width of the flux tube 
(typically proportional to $T_g$) has to be considered. 

\vskip 1truecm
\leftline{\bf Exercises}
\begin{itemize}
\item[6.4.1]{ Show that in the limit $T_g \to 0$ the Wilson loop in  Eq. (\ref{Wsvm}) 
is given by $\exp \{-\sigma S\}$ with $\sigma$ as defined above and $S$ the surface enclosed by the loop $\Gamma$.} 
\item[6.4.2]{ Show that using the lattice parameterization of $D$ given above, with  $T_g \simeq 0.3$ fm and 
$F_2 \simeq 0.048$ GeV$^4$, the Stochastic Vacuum Model gives a string tension $\sigma$ compatible 
with the phenomenological value of Eq. (\ref{sigma1}).}  
\end{itemize}

\subsection{String description of QCD}
An effective string theory of strong interaction has been proposed first by Nambu and Goto in 1960 
to explain the fact that the hadrons lie on Regge trajectories (see Sec. 2). Nambu was 
the first who explicitly constructed an effective theory of the Abelian Higgs model working 
in the limit of infinite Higgs mass which turned out to be coincident with the Nambu--Goto
string Lagrangian. Recent works have shown that properly summing over all the string positions 
to account for the fluctuation of the string generates an effective string in four dimension 
free of the conformal anomaly. Results exist both in the limit of infinite Higgs mass (Akhmedov et al. (1996)) 
and without taking that singular limit (Baker et al. (1999)). A phenomenological description of hybrids has been 
obtained using a Nambu--Goto action in Allen et al. (1998).

\section{Summary}
These lectures have been devoted to a summary of our understanding of the heavy hadron spectrum in QCD. 
We started from some phenomenological guess of the  interquark static potential 
inspired  from a naive flux tube picture. Then, we managed to give a well founded definition of such 
a potential in field theory and to ``derive'' the quark confinement property in  strong 
coupling expansion. To obtain the form of the potential we resorted to the  lattice formulation 
and to numerical techniques that eventually confirmed the linear rising 
of the  phenomenological static potential. The relativistic corrections were added in the framework of effective 
field theories establishing a systematic way of estimating the neglected terms. This allowed us to obtain 
a gauge-invariant field theory based definition of the semirelativistic quark interaction devised  for lattice 
simulations as well as for analytic calculations. Eventually this procedure enabled us to connect directly 
the QCD vacuum with the spectrum. At the end we examined the Wilson loop in the Abelian projection suggested by 't Hooft, 
i.e. under the assumption that the QCD vacuum is a dual superconductor, and we found that our initial 
Cornell potential seems to be the output of dominant monopoles configurations.

The aim of these lectures has been to show how heavy quark bound states offer an ideal situation 
where exact results from QCD (effective field theories like NRQCD and pNRQCD) can be used 
to explore the QCD vacuum either with lattice numerical tools or by means of analytic models. 
Therefore many theoretical ideas and techniques can be and have been used in order to explain 
that kind of systems. All of them cooperate to enlarge our predictability and to open new perspectives. 
It is not surprising that, in spite of several still open problems, some remarkable progress have been achieved.

\section*{Acknowledgments}
We gratefully acknowledge interesting discussions with Marshall Baker, Manfried Faber, Dieter Gromes, Khin Maung, 
Martin Zach and the participants at the HUGS'98 School. We thank Gunnar Bali for making available to us his  lattice data, 
for many enlightening discussions and for making many useful comments and suggestions especially on the lattice part of these 
lectures. We acknowledge M. Baker and  D. Gromes for reading the manuscript and  making useful comments.
It is a pleasure for N. B. to thank  Jos\'e Goity for the invitation to give these lectures 
and for the perfect organization of the school. N. B. acknowledges the support of the European Community, 
Marie Curie fellowship, TMR  contract No. ERBFMBICT961714; A. V. acknowledges the FWF contract No. 9013.

\end{document}